# Generalization of the Fortuin-Kasteleyn transformation and its application to quantum spin simulations

N. Kawashima and J.E. Gubernatis
*Center for Nonlinear Studies and Theoretical Division*
*Los Alamos National Laboratory, Los Alamos, NM 87545*
(February 15, 1995)

## Abstract

We generalize the Fortuin-Kasteleyn (FK) cluster representation of the partition function of the Ising model to represent the partition function of quantum spin models with an arbitrary spin magnitude in arbitrary dimensions. This generalized representation enables us to develop a new cluster algorithm for the simulation of quantum spin systems by the worldline Monte Carlo method. Because the Swendsen-Wang algorithm is based on the FK representation, the new cluster algorithm naturally includes it as a special case. As well as the general description of the new representation, we present an illustration of our new algorithm for some special interesting cases: the Ising model, the antiferromagnetic Heisenberg model with $S = 1$, and a general Heisenberg model. The new algorithm is applicable to models with any range of the exchange interaction, any lattice geometry, and any dimensions.

Typeset using REVTEX

# I. INTRODUCTION

In 1987, Swendsen and Wang [1] used the Fortuin and Kasteleyn (FK) representation [2] of the partition function of Ising models in a Monte Carlo simulation of these models. With the use of this representation, they were able to produce an algorithm whose key feature was the global updating of Ising spin configurations in contrast to local updating in the standard Metropolis algorithm. With global updating, their cluster algorithm greatly reduced the autocorrelation times in the simulation near a critical point. Since then, several attempts have been made [3] to reduce autocorrelation times of various systems by various forms of cluster algorithms, and recently the construction of a cluster algorithm has been formulated on more general grounds [4–7]. Still, most applications of cluster algorithms have been restricted to classical models.

In this paper, we discuss cluster algorithms for worldline Monte Carlo (WLMC) simulations of general classes of quantum spin systems. For most quantum spin systems, exact knowledge of the properties of the systems is very restricted. To obtain information of a wider range, one often resorts to numerical methods such as exact diagonalization, expansions with respect to small parameters, and quantum Monte Carlo methods. Among these, only quantum Monte Carlo methods are available for relatively large systems. Several variants of the quantum Monte Carlo simulation exist: Green's function Monte Carlo (GFMC) method [8], the projector Monte Carlo method [9], the auxiliary-field method based on Hubbard-Stratonovich transformation [10], Handscomb's method [11], and the worldline Monte Carlo (WLMC) method based on the Suzuki-Trotter (ST) approximation [12]. For the study of ground state properties, the GFMC is particularly useful. The auxiliary-field method is powerful for Hubbard models and related problems. Handscomb's method lacks the systematic error caused by the ST approximation. The WLMC has enjoyed a wide range of applicability to fermion, boson, and quantum spin models. For various models, it is the simplest and perhaps the most widely known.

For fermion problems and frustrated spin systems, the main difficulty shared by all the above-mentioned methods is the well-known negative sign problem. In some special cases, one of the methods can be particularly useful compared to others, as is the case for the auxiliary field method applied to the Hubbard model with particle-hole symmetry. However, for many other models, such as frustrated spin models, the WLMC is commonly used because no other method is particularly efficient in reducing the difficulty. In this paper, we do not address the sign problem. We will rather focus on another difficulty which has been under-appreciated. The WLMC suffers from long autocorrelation times even when the system is not near a finite-temperature critical point [13]. Although a similar problem may exist for the other quantum Monte Carlo techniques, this problem has not yet been studied systematically. One way to overcome this difficulty is to develop a cluster updating as in the SW method. In this paper we present the details of such a method.

In general, it is non-trivial to find a cluster algorithm. The generalized approaches [4–7], however, provide a starting point. They first require the specification of a proper set of local graphs by which the whole system is decomposed into clusters and also require a non-negative solution to a system of linear equations (weight equations) that is often under-determined. Little *a priori* guidance is given on the construction of these graphs, and even the existence of a non-negative solution is not guaranteed. Nevertheless, solutions exist in some simple cases. The Swendsen-Wang (SW) algorithm, for instance, is one such case. Here, the number of weight equations is only two as is the number of independent variables. A slightly more complicated case is the loop algorithm [14,15] for the six-vertex model. In the massless case, for example, both the number of equations and the number of variables are three. This algorithm was successfully applied in a WLMC simulation of the spin 1/2 antiferromagnetic Heisenberg model [16] because the $S = 1/2$ quantum spin systems can be mapped to the six-vertex model by using ST approximation [12]. This particular algorithm can also be viewed as the simplest example of the general method that we present in this paper.

For a general quantum spin system, the number of the weight equations and independent variables can be very large. As we will see below, even in the next simplest case, the case of the $XY$-like $XXZ$ model, is already somewhat difficult to handle as the number of the equations is 7 whereas the number of the variables are 11, Here, it does not seem guaranteed that a meaningful solution exists. In fact, however, at least one meaningful solution exists for any system regardless of the magnitude of spins or the coupling constants. This rather surprising result is what we will present in this paper as well as a practical method for obtaining the solution. Instead of working on the complicated weight equation itself, we will take another approach which naturally leads to a proper choice of graphs and a solution of the equations; that is, we will generalize the FK cluster representation of the Boltzmann weight of Ising models to the quantum models. We thereby propose a new algorithm which potentially reduces the autocorrelation times greatly. We have already applied the new algorithm to the $S = 1$ antiferromagnetic Heisenberg chain [17], achieving a reduction of some autocorrelation times by as much as four orders of magnitude. These algorithms differ from previous attempts [18] to generalize the FK transformation in that they produce clusters whose states can naturally be specified by a single variable each and can change independently. In this paper, we present the details of the construction of the algorithm.

It is useful to outline the algorithm before the detailed explanations and the mathematical proofs. The lattice we will work on is not the original lattice on which the quantum problem is defined. Instead, we will consider many lattices, each of which is geometrically equivalent to the original lattice, and take this set of lattices as a hyper-lattice which has dimensions greater than the original dimensions by one. This hyper-lattice can also be viewed as a collection of local units which we call plaquettes and links. One Monte Carlo step of the new algorithm consists of a labeling process applied to the plaquettes and a flipping process for clusters formed by these labels. In the labeling process, we assign a label, i.e., a graph, stochastically to each local unit. More specifically, in each local unit, we connect vertices by edges which are chosen probabilistically. After this graph assignment is done for all plaquettes, the union of all local graphs forms a global graph defined for the entire lattice: the connected vertices form clusters and these clusters constitute the global graph. In the flipping process, each cluster is flipped randomly with probability 1/2. The question to be answered is how do we obtain both the set of local graphs and the labeling probabilities.

In this paper, we will develop various notations and definitions needed to describe our algorithm without ambiguity. To this end, in Section II, we first summarize the general framework of cluster representation and cluster algorithm, and then we illustrate how a particular problem fits into this general framework by presenting two examples, the SW algorithm for the Ising model and the $S = 1$ antiferromagnetic Heisenberg chain. These examples will not only illustrate the notation but also give a peek at the generalization of the FK representation and the Monte Carlo algorithms that we will derive from it. After these examples, formal definitions for the words introduced in the examples will be given, in Section III. In this section, we also present, without derivation, the new cluster representation for the general $XXZ$ models. Section IV is the essential part of this paper. In this section, we will show how to obtain a solution of the weight equation for the general $XXZ$ spin model. We also propose a practical method for computing the solution. In Section V, we present the compact formula for the solution of the weight equation for the Heisenberg models. In Section VI, the ergodicity of the new algorithm is discussed. Section VII is a brief summary. We also discuss a further generalization of our results to the $XYZ$ models in the appendix.

## II. FRAMEWORK AND EXAMPLES

In this section, we first show how the FK-type representation of the partition function is used in a Monte Carlo simulation and then present two pedagogical examples to introduce symbols and notation which we will use in the subsequent sections to generalize the FK representation and to develop the new algorithms.

### A. The cluster representation

In general, the partition function of a lattice spin system is the sum of some weight function over all possible 'spin configurations'

$$Z = \sum_{\bm{n}} W(\bm{n}), \tag{2.1}$$

where

$$\bm{n} \equiv (n_1, n_2, \cdots, n_{N_v}) \tag{2.2}$$

denotes a set of one-bit variables each of which takes a value 0 or 1. This type of configurational summation is the case even with quantum spin systems when they are mapped by the ST decomposition to problems with classical degrees of freedom. Here, we are considering $N_v$ *vertices* which are lattice points in the case of Ising model but will not necessarily be lattice points in the more general cases to be discussed below. In the subsequent sections, we will show that this weight function $W(\bm{n})$ itself can be expressed as a sum of another weight function over a variable different from the 'spin configurations' $\bm{n}$,

$$W(\bm{n}) = \sum_{G} \tilde{W}(\bm{n}, G). \tag{2.3}$$

This new variable $G$ is a graph defined on the lattice. The equation (2.1) with (2.3) can be viewed as a partition function of a system that consists of vertex variables $\bm{n}$ and graphs $G$ interacting each other. A graph consists of *edges* each of which connects two vertices. It imposes a strong restriction on the values that the variable $\bm{n}$ can take because the weight function $\tilde{W}(\boldsymbol{n}, G)$ will be zero for many 'spin configurations'. Therefore, we can express the weight function $\tilde{W}(\boldsymbol{n}, G)$ as

$$\tilde{W}(\boldsymbol{n}, G) = V(G)\Delta(\boldsymbol{n}, G), \tag{2.4}$$

where the function $\Delta(\boldsymbol{n}, G)$ represents the restriction imposed by $G$, i.e., it takes the value 1 if $\boldsymbol{n}$ is *compatible* with $G$ and 0 otherwise.

All the functions that appeared so far also factorize into products where each factor is defined on a local *unit* denoted by $u$, i.e.,

$$W(\boldsymbol{n}) = \prod_u w(\boldsymbol{n}(u)), \tag{2.5}$$

$$V(G) = \prod_u v(G(u)), \tag{2.6}$$

$$\Delta(\boldsymbol{n}, G) = \prod_u \Delta(\boldsymbol{n}(u), G(u)). \tag{2.7}$$

A local unit is a bond, i.e. a pair of vertices, in the case of the SW algorithm, whereas it is more than two vertices in general. The symbol $\boldsymbol{n}(u)$ is the part of $\boldsymbol{n}$ that concerns the unit $u$. The symbol $G(u)$ has the similar meaning. The last equation (2.7) means that the restriction imposed by the graph $G$ is collection of local restrictions imposed by its subgraphs $G(u)$. Re-expressing (2.3) in the factorized form, we obtain

$$w(\boldsymbol{n}(u)) = \sum_{G(u) \in \Gamma(u)} v(G(u))\Delta(\boldsymbol{n}(u), G(u)), \tag{2.8}$$

where $\Gamma(u)$ is the set of graphs for the problem at hand.

We must assume another property of the restriction imposed by $G$: when we define a one-bit cluster variable $\sigma_c = 0$ or 1 on each cluster in $G$, a way must exist for specifying an arbitrary state compatible with $G$ in such a way that the $n_i$ depends on $i$ only through $\sigma_c$ where $c$ is the cluster to which $i$ belongs. Here, a cluster is a maximal group of vertices connected by edges to which another vertex cannot be added without loss of connectivity. In general, a graph $G$ has many clusters. To be more specific, defining $N_c(G)$ as the number of clusters in $G$ and $\Sigma(G)$ as a set of configurations compatible with $G$, we assume a bijection exists which maps $\{0, 1\}^{N_c(G)}$ onto $\Sigma(G)$ in such a way that $(\sigma_1, \sigma_2, \cdots, \sigma_{N_c})$ maps to $(n_1, n_2, \cdots, n_{N_v})$ with $n_i$ depending only on $\sigma_c$ and $i \in c$.

Thus, our problem is clearly defined. Of course, it is not obvious *a priori* that a given partition function can be expressed in the form described here. Most of this paper will be dedicated to the derivation of this type of representation for quantum spin systems. In the next subsection, we will discuss how we can construct a proper Markov process assuming that we are already given the above representation.

### B. Cluster Monte Carlo method

We consider a Markov process $(\boldsymbol{n}^{(t)}, G^{(t)})$ $(t = 1, 2, 3, \cdots)$ on the extended phase space $\Sigma \times \Gamma$ [7], where $\Sigma$ is the configuration space, i.e. the set of possible values that $\boldsymbol{n}$ can take, and $\Gamma$ is the space of graphs. The Markov process is characterized by two transition matrices:

the first one $T_\mathrm{L}(G'|\boldsymbol{n},G)$ gives the probability of having a graph $G'$ given a state $(\boldsymbol{n},G)$, and the second one $T_\mathrm{F}(\boldsymbol{n}'|\boldsymbol{n},G)$ gives the probability of having a new set of vertex variables $\boldsymbol{n}'$. More specifically, $T_\mathrm{L}$ and $T_\mathrm{F}$, called *labeling* probability and the *flipping* probability, are defined by

$$T_\mathrm{L}(G'|\boldsymbol{n},G) \equiv \Pr(G^{(t+1)} = G'|\boldsymbol{n}^{(t)} = \boldsymbol{n}, G^{(t)} = G),$$
$$T_\mathrm{F}(\boldsymbol{n}'|\boldsymbol{n},G) \equiv \Pr(\boldsymbol{n}^{(t+1)} = \boldsymbol{n}'|\boldsymbol{n}^{(t)} = \boldsymbol{n}, G^{(t+1)} = G),$$

where $\Pr(X|Y)$ is the conditional probability of having an event $X$ given an event $Y$.

In [7], a special class of cluster algorithm is discussed in which each cluster can be flipped independently as if it were a single non-interacting degree of freedom. This type of cluster algorithm is called *free*. To obtain a free cluster algorithm [7], we need to choose the flipping probability as

$$T_\mathrm{F}(\boldsymbol{n}'|\boldsymbol{n},G) = \Delta(\boldsymbol{n}',G)/N(G), \tag{2.9}$$

where $N(G) \equiv |\Sigma(G)| = \sum_{\boldsymbol{n}} \Delta(\boldsymbol{n},G)$. As mentioned in the last subsection, the symbol $\Sigma(G)$ stands for a set of configurations defined by

$$\Sigma(G) \equiv \{\boldsymbol{n} \in \Sigma | \Delta(\boldsymbol{n},G) = 1\}. \tag{2.10}$$

Because of the property of $\Delta(\boldsymbol{n},G)$ discussed in the last section, all variables in a cluster are 'locked' into a single one-bit degree of freedom and $N(G)$ equals $2^{N_c(G)}$, where $N_c(G)$ is the total number of clusters.

Therefore, the implication of the specific form (2.9) for the flipping probability is that we flip each cluster in the given graph $G$ at random with probability $1/2$ as if each were a single one-bit degree of freedom not interacting with the others. Although we could choose the flipping probability different from (2.9), the choice we made is obviously advantageous in some respect. Computational simplicity is such an advantage. In addition, we can use the improved estimator [19] for the magnetic cumulants.

Given the flipping probability (2.9), we need to choose the labeling probability as

$$T_\mathrm{L}(G'|\boldsymbol{n},G) = V(G')\Delta(\boldsymbol{n},G')/W(\boldsymbol{n}) \tag{2.11}$$

in order that the limiting distribution is the one desired. It is easy to see that the two transition probabilities (2.9) and (2.11) have the distribution $\tilde{W}(\boldsymbol{n},G)$ as their stationary distribution. In other words, the distribution $\tilde{W}(\boldsymbol{n},G)$ is an eigenstate with the eigenvalue 1 of both the transition matrices, i.e.,

$$\tilde{W}(\boldsymbol{n}',G) = \sum_{\boldsymbol{n} \in \Sigma} T_\mathrm{F}(\boldsymbol{n}'|G)\tilde{W}(\boldsymbol{n},G), \tag{2.12}$$

$$\tilde{W}(\boldsymbol{n},G') = \sum_{G \in \Gamma} T_\mathrm{L}(G'|\boldsymbol{n})\tilde{W}(\boldsymbol{n},G). \tag{2.13}$$

Here we used the fact that $T_\mathrm{F}(\boldsymbol{n}'|\boldsymbol{n},G)$ and $T_\mathrm{L}(G'|\boldsymbol{n},G)$ are independent of $\boldsymbol{n}$ and $G$, respectively, and used the abbreviated notation $T_F(\boldsymbol{n}|G)$ and $T_L(G|\boldsymbol{n})$ for the transition probabilities. Therefore, when the ergodicity holds for the algorithm, the only possible limiting probability-distribution is proportional to $\tilde{W}(\boldsymbol{n},G)$. Indeed, we can prove that the algorithm is ergodic in the case of ferromagnetic and $XY$-like models. In the case of antiferromagnetic model, although we have not proved the ergodicity itself, we can prove that if the conventional algorithm is ergodic, the new algorithm is also ergodic. This issue is discussed in Section VI.

We also note that because of (2.3)

$$\sum_{G \in \Gamma} \tilde{W}(\boldsymbol{n}, G) \tag{2.14}$$

is the distribution that we want to generate. This fact implies that we can obtain the distribution $W(\boldsymbol{n})$ simply by generating a Markov sequence described above, picking the vertex-variable part $\boldsymbol{n}^{(t)}$ from the state $(\boldsymbol{n}^{(t)}, G^{(t)})$, and ignoring the graph part $G^{(t)}$.

In an actual Monte Carlo simulation, the labeling process is done locally because of the decomposability of $V(G)\Delta(\boldsymbol{n}, G)$, i.e., (2.6) and (2.7). Therefore, we can generate a graph $G$ with the probability (2.11), simply by picking a graph $G(u)$ for each local unit with probability

$$P_{\rm L}(G(u)|\boldsymbol{n}(u)) \equiv v(G(u))\Delta(\boldsymbol{n}(u), G(u))/w(\boldsymbol{n}(u)), \tag{2.15}$$

and then taking the union of these $G(u)$'s as $G$. We can easily see that the probability (2.15) is properly normalized because of (2.8).

### C. The Fortuin-Kasteleyn representation and the Swendsen-Wang algorithm

In this subsection, we briefly review the simplest and best-known cluster algorithm, the SW algorithm for the Ising model, and its connection with the FK representation. This algorithm and connection have already been discussed [3]; however, we need a more formal discussion to achieve our goal of establishing a mathematically rigorous background for general cluster algorithms. We also found that Fortuin and Kasteleyn's definitions and notation are not completely convenient to describe more general and complicated algorithms. Therefore, in this subsection, we will review the FK representation and the SW algorithm in a language which will eventually develop to accommodate more general ideas as we proceed.

The Hamiltonian of the Ising model can be written as

$$\mathcal{H} = -J \sum_{(i,j)} (2n_i - 1)(2n_j - 1), \tag{2.16}$$

where $n_i = 0, 1$. Correspondingly, the partition function of the Ising model is written as

$$Z = \sum_{\boldsymbol{n}} W(\boldsymbol{n}). \tag{2.17}$$

Here $\boldsymbol{n} \equiv (n_1, n_2, \cdots, n_{|L|})$ is a set of $|L|$ one-bit variables where $|L|$ is the number of lattice points of the lattice $L$. The function $W(\boldsymbol{n})$ is the Boltzmann weight which can be decomposed into a product of local factors defined on local units, i.e., bonds in this case,

$$W(\boldsymbol{n}) = \prod_{b} w(\boldsymbol{n}(b)), \tag{2.18}$$

where $\boldsymbol{n}(b) \equiv (n_i, n_j)$ is a part of $\boldsymbol{n}$ concerning the two end-points $i$ and $j$ of a bond $b$. The local Boltzmann factor $w(\boldsymbol{n}(b))$ is then defined by

$$w(\boldsymbol{n}(b)) \equiv \exp(K(2n_i - 1)(2n_j - 1)), \tag{2.19}$$

where $K \equiv \beta J$. Equation (2.19) can be expressed as a sum of two terms each of which corresponds to one of two graphs often called 'deleted' and 'frozen'

$$w(\boldsymbol{n}(b)) = \sum_{g=d,f} v(g) \Delta(\boldsymbol{n}(b), g), \tag{2.20}$$

where

$$v(d) = e^{-K} \quad \text{and} \quad \Delta(\boldsymbol{n}(b), d) = 1 \tag{2.21}$$

for a 'deleted' bond and

$$v(f) = e^{K} - e^{-K} \quad \text{and} \quad \Delta(\boldsymbol{n}(b), f) = \delta(n_i, n_j) \tag{2.22}$$

for a 'frozen' bond. Substituting (2.20) into (2.18), we have

$$W(\boldsymbol{n}) = \prod_b \sum_{g_b} v(g_b) \Delta(\boldsymbol{n}(b), g_b),$$
$$= \sum_{\{g_b\}} (\prod_b v(g_b))(\prod_b \Delta(\boldsymbol{n}(b), g_b)), \tag{2.23}$$

The product $\prod_b \Delta(\boldsymbol{n}(b), g_b)$ implies that a configuration $\boldsymbol{n}$ is allowed only if two variables $n_i$ and $n_j$ at the end-points of any frozen bond are equal. Otherwise, this product vanishes and the corresponding term does not contribute to the sum. We can visualize this situation by placing edges on frozen bonds and nothing on deleted bonds. These edges form a graph which we denote as $G$. Obviously, this graph has the same information as the set of variables $\{g_b\}$. Therefore, we will identify these two and simply write $G = \{g_b\}$. A cluster is a maximal set of vertices connected to each other by edges. A graph consists of many clusters, in general. It is clear that for a given graph $G$ an allowed state is one where all variables $n_i$ in the same cluster in $G$ have the same value 0 or 1. In other words, all variables in a cluster are locked into a single degree of freedom represented by $\sigma_c = 0$ or 1 where $c$ specifies a cluster in $G$.

If we adopt the notation used in (2.7), we can write $g_b = G(b)$, and then rewrite (2.23) as

$$W(\boldsymbol{n}) = \sum_G V(G) \Delta(\boldsymbol{n}, G), \tag{2.24}$$

where

$$V(G) \equiv \prod_b v(G(b)), \tag{2.25}$$

$$\Delta(\boldsymbol{n}, G) \equiv \prod_b \Delta(\boldsymbol{n}(b), G(b)). \tag{2.26}$$

$$\tag{2.27}$$

Substituting (2.24) into (2.17), we have

$$Z = \sum_{\boldsymbol{n}} \sum_{G} V(G) \Delta(\boldsymbol{n}, G). \tag{2.28}$$

This equation is the Fortuin-Kasteleyn representation for the Ising model. Although it is not explicitly indicated, $Z$ and $V(G)$ depend on the model and the coupling constant $K$. On the other hand, $\Delta(\boldsymbol{n}, G)$ will be used for other models by generalizing its definition. In what follows, we will see exactly the same form as (2.28) with different $Z$ and $V(G)$ for other models.

Because we have rewritten the partition function in the form discussed in Subsection II A, we can construct the Markov process following the general prescription given in Subsection II B. The resulting algorithm is the SW algorithm. To be specific, given a configuration $\boldsymbol{n}$, we first choose a graph $G$ with the weight $V(G)\Delta(\boldsymbol{n}, G)$ and then pick a new configuration $\boldsymbol{n}'$ with the weight $\Delta(\boldsymbol{n}', G)$. The first step is equivalent to choosing a local graph $G(b)$ with the weight $v(G(b))\Delta(\boldsymbol{n}(b), G(b))$. Because of (2.20), the probability of assigning $G(b)$ to a given bond in state $\boldsymbol{n}(b)$ is

$$P_L(G(b)|\boldsymbol{n}(b)) \equiv v(G(b))\Delta(\boldsymbol{n}(b), G(b))/w(\boldsymbol{n}(b)). \tag{2.29}$$

This result agrees with the ordinary SW labeling probability

$$P_L(\text{`deleted'}|(0,0)) = P_L(\text{`deleted'}|(1,1)) = e^{-2K}, \tag{2.30}$$

$$P_L(\text{`deleted'}|(0,1)) = P_L(\text{`deleted'}|(1,0)) = 1, \tag{2.31}$$

$$P_L(\text{`frozen'}|\boldsymbol{n}(b)) = 1 - P_L(\text{`deleted'}|\boldsymbol{n}(b)). \tag{2.32}$$

On the other hand, the second step is, as discussed in the last subsection, equivalent to setting each cluster variable $\sigma_c$ to 0 or 1 with equal probability.

### D. Generalization to quantum spin systems

In this subsection, we will give an example of the generalization of the FK representation to quantum spin systems. We will discuss the $S = 1$ antiferromagnetic Heisenberg model in one dimension. The Hamiltonian is written as

$$\mathcal{H} = J \sum_{i=1}^{N} \boldsymbol{S}_i \cdot \boldsymbol{S}_{i+1}, \tag{2.33}$$

where $\boldsymbol{S}_i^2 = 2$ and

$$[S_i^{\alpha}, S_j^{\beta}] = \sqrt{-1}\delta_{i,j}S_j^{\gamma}, \qquad (\alpha, \beta, \gamma) = (x, y, z), (y, z, x), \text{ or } (z, x, y). \tag{2.34}$$

The periodic boundary condition, i.e., $\boldsymbol{S}_{N+1} = \boldsymbol{S}_1$ is assumed. We also assume that $N$ is even. Then, by applying the unitary transformation

$$S_i^x \to -S_i^x \quad \text{and} \quad S_i^y \to -S_i^y \tag{2.35}$$

to all sites with even $i$, we have

$$\mathcal{H} = -J \sum_{i=1}^{N} \hat{\Lambda}_{i,i+1}, \tag{2.36}$$

where

$$\hat{\Lambda}_{i,j} \equiv S_i^x S_j^x + S_i^y S_j^y - S_i^z S_j^z. \tag{2.37}$$

We remark that these assumptions are not essential to the general algorithm described in the following sections.

Usually, we take the basis in which $S_i^z$ is diagonalized for Monte Carlo simulations. Then, a basis vector is specified by a set of $N$ classical variables, each of which takes a value $-1, 0$ or $1$. The Hilbert space is, of course, $3^N$ dimensional. However, this basis is inconvenient for our purpose of mapping the problem into a one-bit problem of the form discussed in Subsection II A. Therefore, we will first map the original quantum problem into another quantum problem with one-bit degrees of freedom. To do this, we first decompose each spin operator into a sum of two Pauli operators.

$$S_i^\alpha = \frac{1}{2}(\sigma_{(i,1)}^\alpha + \sigma_{(i,2)}^\alpha), \qquad \alpha = x, y, \text{ or } z, \tag{2.38}$$

where

$$[\sigma_{(i,\mu)}^\alpha, \sigma_{(j,\nu)}^\beta] = 2\sqrt{-1}\delta_{i,j}\delta_{\mu,\nu}\sigma_{(i,\mu)}^\gamma. \tag{2.39}$$

Then, we take a basis in which the $z$-components of the Pauli operators are diagonalized. Therefore, a basis vector can be expressed as $|n\rangle$ where $n$ is a set of $2N$ one-bit variables

$$\boldsymbol{n} = (n_{(1,1)}, n_{(1,2)}, n_{(2,1)}, n_{(2,2)}, n_{(3,1)}, n_{(3,2)}, \cdots n_{(N,1)}, n_{(N,2)}), \tag{2.40}$$

and has the property

$$\sigma_{(i,\mu)}^z |n\rangle = (2n_{(i,\mu)} - 1)|n\rangle, \qquad n_{(i,\mu)} = 0 \text{ or } 1. \tag{2.41}$$

Clearly, the Hilbert space is $2^{2N}$ dimensional and larger than the original Hilbert space. The new Hilbert space is larger because it includes some singlet states which are unphysical here. Since we should not count such states in the partition function, the partition function in this basis becomes

$$Z = \sum_{\boldsymbol{n}} \langle \boldsymbol{n} | P e^{-\beta \mathcal{H}} P | \boldsymbol{n} \rangle. \tag{2.42}$$

The projection operator $P$ is the product of local projection operators $P_i$ which projects out states with $\boldsymbol{S}_i^2 = 3/4$

$$P = \prod_i P_i. \tag{2.43}$$

It is easy to see that $P_i$ is the symmetrization operator

$$P_i = \frac{1}{2}(\hat{I} + \hat{X}_i) \tag{2.44}$$

where $\hat{I}$ is the identity operator and $\hat{X}_i$ is the operator that exchanges $n_{(i,1)}$ and $n_{(i,2)}$.

Thus, the original problem is mapped to a problem with one-bit degrees of freedom (2.42). This problem still is a quantum problem which needs the evaluation of matrix elements of the Boltzmann density operator. Therefore, the next thing we must do is to map this problem to a classical one-bit problem by the ST decomposition. We can write $Pe^{-\beta\mathcal{H}}P$ in (2.42) as

$$Pe^{-\beta\mathcal{H}}P = (Pe^{-\Delta\tau\mathcal{H}}P)^{M_T} \approx [(Pe^{-\Delta\tau\mathcal{H}_\mathrm{E}}P)(Pe^{-\Delta\tau\mathcal{H}_\mathrm{O}}P)]^{M_T}, \tag{2.45}$$

where $M_T\Delta\tau \equiv \beta$ and

$$\mathcal{H}_\mathrm{E} \equiv -J\sum_{i:even}\hat{\Lambda}_{i,i+1}, \quad \mathcal{H}_\mathrm{O} \equiv -J\sum_{i:odd}\hat{\Lambda}_{i,i+1}. \tag{2.46}$$

The integer $M_T$ is called Trotter-number. By inserting the identity operator

$$I = \sum_{\bm{n}'}|\bm{n}'\rangle\langle\bm{n}'| \tag{2.47}$$

between the factors in (2.45), we can re-express the partition function (2.42) as

$$Z = \sum_{\bm{n}_M}\sum_{\bm{n}_{M-1}}\cdots\sum_{\bm{n}_1}$$
$$\langle\bm{n}_1|Pe^{-\Delta\tau\mathcal{H}_\mathrm{E}}P|\bm{n}_M\rangle\langle\bm{n}_M|Pe^{-\Delta\tau\mathcal{H}_\mathrm{O}}P|\bm{n}_{M-1}\rangle\cdots\langle\bm{n}_2|Pe^{-\Delta\tau\mathcal{H}_\mathrm{O}}P|\bm{n}_1\rangle, \tag{2.48}$$

where $M \equiv 2M_T$.

Because of the form of (2.48), it is natural to introduce a hyper-lattice which consists of $M$ lattices, each of which is equivalent to the original lattice. Because the original lattice is a ring in the present case and we are automatically imposing the periodic boundary condition in the new direction, the hyper-lattice is a torus. We will refer to this hyper-lattice by a symbol $\tilde{L}$ and each ring in this hyper-lattice by a symbol $\tilde{L}_k$ ($k = 1, 2, \cdots, M$). We call $MN$ lattice points in the hyper-lattice *sites*. As we have seen, two one-bit variables are defined on each site. They can be viewed as variables defined on two *vertices* inside the site. Thus, the hyper-lattice is a set of $2MN$ vertices. In what follows, we refer to the set of vectors $\{\bm{n}_1, \bm{n}_2, \cdots, \bm{n}_M\}$ simply as $\bm{n}$. We also refer to the variable $n_{(i,\mu)}$ on $\tilde{L}_k$ as $n_{(k,i,\mu)}$. Accordingly, the vertex on which $n_{(k,i,\mu)}$ is defined will be referred to as $(k,i,\mu)$. The site that contains the vertices $(k,i,1)$ and $(k,i,2)$ will be referred to as $(k,i)$. In Fig. 1, the hyper-lattice is shown for the case of $N = 6$ and $M_T = 2$. Next, we introduce the symbol $\bm{n}(V)$ that stands for the part of $\bm{n}$ concerning a set $V$ where $V$ is any set of vertices. For example, $\bm{n}((k,i,\mu))$ is simply $n_{(k,i,\mu)}$, $\bm{n}((k,i))$ stands for $(n_{(k,i,1)}, n_{(k,i,2)})$, $\bm{n}(\tilde{L}_k)$ is $\bm{n}_k$, and $\bm{n}(L)$ is $\bm{n}$ itself. Actually, a special case of this notation has already been used in (2.7).

Having defined this notation, we note that the state $|\bm{n}(\tilde{L}_k)\rangle$ can be expressed by a direct product of local wave functions as

$$|\bm{n}(\tilde{L}_k)\rangle = \bigotimes_{i:even}|\bm{n}(l_{(k,i)})\rangle_2, \tag{2.49}$$

where $l_{(k,i)}$ is a pair of sites $\{(k,i), (k,i+1)\}$, or equivalently a set of four vertices $\{(k,i,1), (k,i,2), (k,i+1,1), (k,i+1,2)\}$.

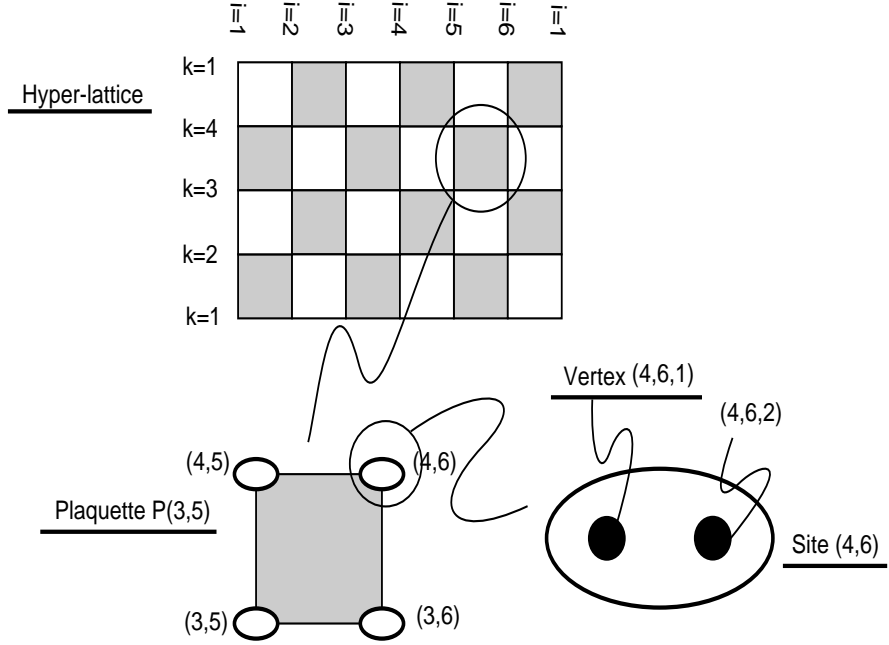

FIG. 1. The 'checkerboard' hyper-lattice for the $S = 1$ $XY$ model on a chain of the length 6 with Trotter number 2. Only shaded plaquettes are called 'plaquettes' here. On each plaquette four-body Boltzmann factor is defined.

To be more specific, $|n(l_{(k,i)})\rangle_2$ is an eigenvector of four Pauli operators $\sigma^z_{(i,\mu)}$ and $\sigma^z_{(i+1,\mu)}$ ($\mu = 1, 2$) in a $2^4$ dimensional local Hilbert space. Therefore, because $[\hat{\Lambda}_{i,i+1}, \hat{\Lambda}_{j,j+1}] = 0$ if $|i - j| \geq 2$, we obtain

$$\langle n(\tilde{L}_{k+1})|Pe^{-\Delta\tau \mathcal{H}_\mathrm{E}} P|n(\tilde{L}_k)\rangle = \prod_{i:even} w(n(p_{(k,i)})), \quad (2.50)$$

where $p_{(k,i)}$ is a *plaquette* defined by $p_{(k,i)} \equiv l_{(k,i)} \cup l_{(k+1,i)}$, and

$$w(n(p_{(k,i)})) \equiv \langle n(l_{(k+1,i)})|P_i P_{i+1} e^{K\hat{\Lambda}_{i,i+1}} P_i P_{i+1}|n(l_{(k,i)})\rangle_2, \quad (2.51)$$

with $K \equiv \Delta\tau J$. An expression similar to (2.50) is available for $\mathcal{H}_\mathrm{O}$. Therefore, the partition function (2.48) can be rewritten as

$$Z = \sum_{n} W(n), \quad (2.52)$$

where

$$W(n) = \prod_{p} w(n(p)). \quad (2.53)$$

The product in (2.53) is taken over $p_{(k,i)}$ with even $k+i$.

Here we note that $w(\boldsymbol{n}(p))$ depends $\boldsymbol{n}(p)$ only through $m_{(k,i)} \equiv \sum_\mu n_{(k,i,\mu)}$ where $(k,i)$ is one of four sites in the plaquette $p$. Therefore, we can re-express $w(\boldsymbol{n}(p))$ in the form

$$w(\boldsymbol{n}(p)) = \tilde{w}\begin{pmatrix} m_{(k+1,i)} & m_{(k+1,i+1)} \\ m_{(k,i)} & m_{(k,i+1)} \end{pmatrix}. \tag{2.54}$$

We further note that this $\tilde{w}$ has several symmetry properties

$$\tilde{w}\begin{pmatrix} m_{\rm tl} & m_{\rm tr} \\ m_{\rm bl} & m_{\rm br} \end{pmatrix} = \tilde{w}\begin{pmatrix} m_{\rm bl} & m_{\rm br} \\ m_{\rm tl} & m_{\rm tr} \end{pmatrix}$$
$$= \tilde{w}\begin{pmatrix} m_{\rm tr} & m_{\rm tl} \\ m_{\rm br} & m_{\rm bl} \end{pmatrix} = \tilde{w}\begin{pmatrix} 2S - m_{\rm bl} & 2S - m_{\rm br} \\ 2S - m_{\rm tl} & 2S - m_{\rm tr} \end{pmatrix} \tag{2.55}$$

where $S=1$ in the present case and the suffix 'tl' means top-left corner of the plaquette, 'tr' means top-right, etc. These symmetry properties define classes of states which correspond to the same local weight $\tilde{w}$ regardless of the value of $K$. In what follows, we will specify one of these classes by a symbol $\mathcal{S}(p)$. The 'particle number conservation', imposes a restriction on $\tilde{w}$, i.e., $\tilde{w}$ is non-zero only if

$$m_{\rm bl} + m_{\rm br} = m_{\rm tl} + m_{\rm tr}. \tag{2.56}$$

We can eliminate from consideration some of the classes that violate this condition. As a result, only seven distinct values among $2^8 = 256$ values of $w(\boldsymbol{n}(p))$ exist, i.e., there are seven relevant classes of states,

$$\begin{aligned}
\tilde{w}(1) &= \tilde{w}\begin{pmatrix} 0 & 0 \\ 0 & 0 \end{pmatrix} = 12r, \\
\tilde{w}(2) &= \tilde{w}\begin{pmatrix} 0 & 2 \\ 0 & 2 \end{pmatrix} = 2r + 6r^{-1} + 4r^{-2}, \\
\tilde{w}(3) &= \tilde{w}\begin{pmatrix} 2 & 0 \\ 0 & 2 \end{pmatrix} = 2r - 6r^{-1} + 4r^{-2}, \\
\tilde{w}(4) &= \tilde{w}\begin{pmatrix} 1 & 1 \\ 0 & 2 \end{pmatrix} = -4r + 4r^{-2}, \\
\tilde{w}(5) &= \tilde{w}\begin{pmatrix} 1 & 0 \\ 0 & 1 \end{pmatrix} = -6r + 6r^{-1}, \\
\tilde{w}(6) &= \tilde{w}\begin{pmatrix} 0 & 1 \\ 0 & 1 \end{pmatrix} = 6r + 6r^{-1}, \\
\tilde{w}(7) &= \tilde{w}\begin{pmatrix} 1 & 1 \\ 1 & 1 \end{pmatrix} = 8r + 4r^{-2},
\end{aligned} \tag{2.57}$$

where $r \equiv \exp(-K)$.

Equations (2.52) and (2.53) have the same form as (2.1) and (2.5). Therefore, the next thing we have to find is a set of local graphs $\Gamma(p)$ and coefficients $v(G(p))$ that satisfy (2.8) with $u$ replaced by $p$

$$w(\boldsymbol{n}(p)) = \sum_{G(p) \in \Gamma(p)} v(G(p)) \Delta(\boldsymbol{n}(p), G(p)). \tag{2.58}$$

To this end, we consider the graphs which correspond to pairing of the eight vertices in $p$, i.e., graphs which consist of four edges sharing no vertices. From the 105 such graphs, we take only those graphs that consist of vertical or horizontal edges. Here, a vertical edge connects two vertices whose spatial indices are the same and the temporal indices are different whereas a horizontal edge connects two vertices whose spatial indices are different and the temporal indices are the same. There are 24 such graphs. When we neglect distinctions between vertices in the same site, these 24 graphs are classified into three classes which are represented by the diagrams in the leftmost column in Fig. 2. We use a symbol $\mathcal{G}(p)$ to specify a class of graphs.

| $\mathcal{G}(p) \diagdown \mathcal{S}(p)$ | 1 $\begin{pmatrix} 0 & 0 \\ 0 & 0 \end{pmatrix}$ | 2 $\begin{pmatrix} 0 & 2 \\ 0 & 2 \end{pmatrix}$ | 3 $\begin{pmatrix} 2 & 0 \\ 0 & 2 \end{pmatrix}$ | 4 $\begin{pmatrix} 1 & 1 \\ 0 & 2 \end{pmatrix}$ | 5 $\begin{pmatrix} 1 & 0 \\ 0 & 1 \end{pmatrix}$ | 6 $\begin{pmatrix} 0 & 1 \\ 0 & 1 \end{pmatrix}$ | 7 $\begin{pmatrix} 1 & 1 \\ 1 & 1 \end{pmatrix}$ |
|---|---|---|---|---|---|---|---|
| 1 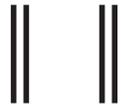 | 4 | 4 | 0 | 0 | 0 | 2 | 1 |
| 2 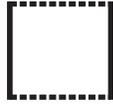 | 0 | 16 | 0 | 4 | 4 | 4 | 2 |
| 3 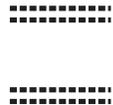 | 0 | 4 | 4 | 2 | 0 | 0 | 1 |

FIG. 2. The coefficient $N(\mathcal{S}(p), \mathcal{G}(p))$ for the $S = 1$ antiferromagnetic Heisenberg model. $\mathcal{S}(p)$ specifies a class of states while $\mathcal{G}(p)$ specifies a class of graphs. In the diagrams of the leftmost column, a solid line stands for a green edge and a dashed line a red edge.

We next define $\Delta(\boldsymbol{n}(p), G(p))$ in (2.58) as the function that takes the value of 1 only if, for any edge connecting two vertices $v$ and $v'$, $\boldsymbol{n}(v) = \boldsymbol{n}(v')$ when the edge is vertical and $\boldsymbol{n}(v) = 1 - \boldsymbol{n}(v')$ when the edge is horizontal. Otherwise, this function takes the value of zero. For later convenience, we call edges for which two vertex variables have the same value *green* and all other edges *red*. In the present case, all vertical edges are green whereas all horizontal edges are red. All edges in what follows have this additional attribute, i.e., 'color'. With this definition, we can formally specify the set of graphs $\Gamma(p)$ in (2.58) as

$$\Gamma(p) \equiv \{G(p) | \text{"Every vertex is an end-point of one and only one edge" and}$$
$$\text{"Any edge is either green vertical or red horizontal."}\} \qquad (2.59)$$

Given (2.57) and (2.59), we next find a solution $v(G(p))$ of (2.58). It is natural to seek a symmetric solution in which $v(G(p))$ depends on $G(p)$ only through the class to which $G(p)$ belongs, i.e., $v(G(p)) = \tilde{v}(\mathcal{G}(p))$ where $G(p) \in \mathcal{G}(p)$. Taking this into account, we can rewrite (2.58) as

$$\tilde{w}(\mathcal{S}(p)) = \sum_{\mathcal{G}(p)} N(\mathcal{S}(p), \mathcal{G}(p)) \tilde{v}(\mathcal{G}(p)). \qquad (2.60)$$

Here, the function $N$ is defined by

$$N(\mathcal{S}(p), \mathcal{G}(p)) \equiv \sum_{G(p) \in \mathcal{G}(p)} \Delta(\boldsymbol{n}(p), G(p)), \qquad (2.61)$$

where $\boldsymbol{n}(p) \in \mathcal{S}(p)$. The function $N$ is shown in Fig. 2. We can easily see that the solution

$$\tilde{v}(1) = \frac{1}{4}\tilde{w}(1),$$
$$\tilde{v}(2) = \frac{1}{4}\tilde{w}(5),$$
$$\tilde{v}(3) = \frac{1}{4}\tilde{w}(3) \qquad (2.62)$$

satisfies (2.60).

Now, since we obtained the representation in the form discussed in Subsection II A, the Monte Carlo method in Subsection II B applies. The resulting algorithm is the loop algorithm used in [17].

In this subsection, we found the solution $v(G(p))$ given a set of graphs $\Gamma(p)$, but did not show how we choose $\Gamma(p)$ or how we obtain the solution $v(G(p))$ in general. These questions will be answered in the following sections.

### III. CLUSTER REPRESENTATION OF GENERAL $XXZ$ MODEL

In Section II D, we saw how $S = 1$ antiferromagnetic Heisenberg chain is mapped to one-bit classical problem. In this section, we will see how the mapping is done in the case of general $XXZ$ spin systems. Our task can be divided into two parts. In Subsection III A, we formulate the problem as a one-bit classical problem with a weight function that factorizes into a product of local factors. Each one of the local factors is defined on a local unit called a plaquette, as we saw already. At this stage, graphs are not yet introduced. In short, in Subsection III A, we will formulate the problem in the form of (2.1) with (2.5). Then, in Subsection III C, we show without derivation how $w(\boldsymbol{n}(u))$ can be expressed as a sum of terms each of which corresponds to a local graph $G(u)$; that is, we will present an explicit form of (2.8) for the $XXY$ model. This expression leads to (2.3). The derivation will be presented in Section IV. To make the discussions clearer, in Subsection III B, we summarize and generalize the notations and definitions introduced in Subsection II D concerning graphs.

## A. Mapping to a classical problem

Our Hamiltonian is

$$\mathcal{H} = -\sum_{(i,j)} (J^x_{i,j} S^x_i S^x_j + J^y_{i,j} S^y_i S^y_j + J^z_{i,j} S^z_i S^z_j)$$
$$= -\sum_{(i,j)} J_{i,j} (S^x_i S^x_j + S^y_i S^y_j + \lambda_{i,j} S^z_i S^z_j), \tag{3.1}$$

with $J^x_{i,j} = J^y_{i,j} = J_{i,j}$ and $J^z_{i,j} = J_{i,j}\lambda_{i,j}$. The operator $S^\alpha_i$ is the spin operator which satisfies

$$\boldsymbol{S}^2_i = (S^x_i)^2 + (S^y_i)^2 + (S^z_i)^2 = S(S+1) \tag{3.2}$$

and

$$[S^\alpha_i, S^\beta_i] = \frac{\sqrt{-1}}{2} S^\gamma_i \tag{3.3}$$

for $(\alpha, \beta, \gamma) = (x, y, z)$, $(y, z, x)$, or $(z, x, y)$. The symbol $(i, j)$ in (3.1) is an arbitrary undirected pair of elements of the set of lattice points $L$. The constants $J_{i,j}$ and $\lambda_{i,j}$ can be any real numbers. We are not assuming any particular geometric feature of the lattice. It can have any dimension and does not even have to be translationally symmetric.

The first important step is to express a spin operator in terms of sum of $2S$ Pauli operators as we did in Section II D,

$$S^\alpha_i = \frac{1}{2} \sum_{\mu=1}^{2S} \sigma^\alpha_{i,\mu}. \tag{3.4}$$

To each Pauli matrix, we will assign a vertex. Therefore, it is convenient to define $\tilde{L}$ as a set of vertices defined on the lattice $L$. We work with a complete set of states that are simultaneous eigenfunctions of the operators $\sigma^z_{i,\mu}$ ($i = 1, 2, \cdots, |L|$; $\mu = 1, 2, \cdots, 2S$). Here, $|L|$ is the number of the lattice points in the lattice $L$. To be more specific,

$$\sigma^z_{i,\mu} |\boldsymbol{n}(\tilde{L})\rangle \equiv (2n_{i,\mu} - 1) |\boldsymbol{n}(\tilde{L})\rangle, \tag{3.5}$$

where $\boldsymbol{n}(\tilde{L})$ represents $2S|L|$ one-bit variables,

$$\boldsymbol{n}(\tilde{L}) \equiv (n_{(1,1)}, n_{(1,2)}, \cdots, n_{(|L|,2S)}). \tag{3.6}$$

The partition function in this basis becomes

$$Z = \sum_{\boldsymbol{n}(\tilde{L})} \langle \boldsymbol{n}(\tilde{L}) | P e^{-\beta \mathcal{H}} P | \boldsymbol{n}(\tilde{L}) \rangle, \tag{3.7}$$

where $P$ is the projection operator defined by

$$P = \prod_{i \in L} P_i, \tag{3.8}$$

and $P_i$ is the projection operator to the space in which $\boldsymbol{S}_i^2$ takes the value $S(S+1)$. Since this representation with Pauli matrices provides a single representation for which $\boldsymbol{S}_i^2 = S(S+1)$, (3.7) is the correct representation of the original model.

As we saw in Subsection II D, using the ST approximation [12] and (3.4), we can map the original quantum problem into a problem with classical one-bit degrees of freedom. However, since there are so many variants of the ST approximation, it is impractical to describe all of them. Therefore, in what follows, we will only describe properties which are shared by all known variants.

Once mapped, the problem has $2SM|L|$ one-bit variables, where $M$ is some integer proportional to the Trotter-number $M_T$. The proportionality constant depends on the variant of the ST approximation. The $2SM|L|$ variables naturally fit into a $d+1$ dimensional hyper-lattice, where $d$ is the dimension of the original lattice $L$. We call the first dimension *temporal* or *vertical* and the other $d$ dimensions which correspond to the dimensions of the original lattice *spatial* or *horizontal*. This lattice consists of $M$ layers of $d$-dimensional lattices each of which is equivalent to the original lattice $L$. We number these layers with an index $k = 1, 2, \cdots, M$ and use $L_k$ to denote the $k$-th layer. We call a lattice point in this hyper-lattice a *site*. $2S$ vertices are associated with each site, and a one-bit variable is defined on each vertex. We label a site with two indices $(k, i)$, with the first index specifying the temporal location of the site, and the second index specifying the spatial location. As for vertices, we use symbols $(k, i, \mu)$ by adding an additional index $\mu = 1, 2, \cdots, 2S$ specifying the $\mu$-th vertex in the site $(k, i)$. The boundary condition in the temporal direction is periodic, i.e., $k = M + 1$ is identified with $k = 1$. On the other hand, we are not assuming any particular spatial geometry; therefore, the present scheme can be applied to models in any dimensions, with any boundary conditions, and with any range of the interactions.

A *horizontal link* is a pair of sites whose temporal indices are the same. A *vertical link* is a pair of sites whose spatial indices are the same and temporal indices differ by one. Some of the squares which consist of two vertical links and two horizontal links play a special role, because a local weight function is defined on them. We call such a square a *plaquette* and often use a symbol $p$ for it. The set of the $8S$ vertices associated with four corners (sites) of a plaquette will also be referred to as a plaquette and represented by the same symbol. Which square among all possible squares is to be a plaquette depends on the variant of the ST approximation. In any variant of the ST approximation, no vertical link is shared by more than one plaquette. The symbol $\mathcal{U}_p$ denotes the set of plaquettes whereas the symbol $\mathcal{U}_l$ denotes the set of vertical links which do not belong to any plaquette. In the example presented in Subsection II D, $\mathcal{U}_l$ is an empty set. In other cases, such as the systems defined on a triangular lattice, $\mathcal{U}_l$ is non-empty. We also define $\mathcal{U} \equiv \mathcal{U}_p \cup \mathcal{U}_l$. This is the set of local units that we will consider in what follows and the products in (2.5), (2.6) and (2.7) should be taken over this set in the present case.

It is convenient to have notations by which we can refer to a specific site or a link in a given plaquette $p$. To this end, we consider a plaquette $p$ consisting of four sites $(k, i), (k, j), (k+1, i)$ and $(k+1, j)$. We first take either one of spatial indices $i$ and $j$ and call it 'left' and the other 'right'. We use the symbols $i_l(p)$ for 'left' index and $i_r(p)$ for the other. (Which one we call 'left' does not matter in what follows.) We also define symbols $l_t(p)$ and $l_b(p)$ as the top and bottom horizontal links of $p$. The symbol $s_{tl}(p)$ stands for a top-left site. Other symbols $s_{tr}(p), s_{bl}(p)$ and $s_{br}(p)$ are defined in a similar fashion. The relative locations of these symbols are illustrated in Fig. 3.

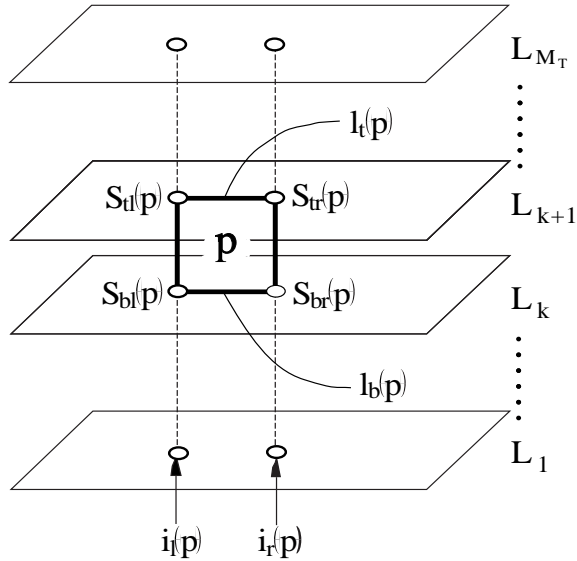

FIG. 3. The hyper-lattice on which the transformed problem is defined and a plaquette on which a four-body Boltzmann factor is defined. The upper link of the plaquette $p$ belongs to the layer $L_{k+1}$ and its lower link belongs to the layer $L_k$. The open ovals represent sites.

With these definitions and the ST approximation, we can rewrite the partition function (3.7) as

$$Z \equiv \sum_{\bm{n} \in \Sigma} \mathrm{sgn}(\bm{n}) W(\bm{n}), \tag{3.9}$$

with the sign factor $\mathrm{sgn}(\bm{n})$ and the Boltzmann weight $W(\bm{n})$ defined below. The sign factor is defined so that $W(\bm{n})$ is always non-negative. The configurations with negative sign appear only for a frustrated system such as the antiferromagnetic Heisenberg model on a triangular lattice. Although the sign factor is physically important, from the computational point of view, it is needed only when adding up measured values at each Monte Carlo step and does not affect the Markov process. In other words, we usually neglect the sign factor in defining the Markov process and the limiting distribution of the resulting process is $W(\bm{n})$. Since the goal of the present paper is to accelerate the Markov process, we neglect the sign factor in (3.9). The weight function $W(\bm{n})$ in (3.9) is defined by

$$W(\bm{n}) \equiv \prod_{p \in \mathcal{U}_p} w(\bm{n}(p)) \prod_{l \in \mathcal{U}_l} \delta_l(\bm{n}(l)). \tag{3.10}$$

Since there is no interaction which corresponds to a vertical link in $\mathcal{U}_l$, $\delta_l(\bm{n}(l))$ should be a function which takes the non-zero value 1 if and only if $n_{(k,i,\mu)} = n_{(k+1,i,\mu)}$ for all $\mu$ where $l = \{(k,i), (k+1,i)\}$.

The function $w(\boldsymbol{n}(p))$ in (3.10) is the absolute value of matrix element of the local Boltzmann operator, that is,

$$w(\boldsymbol{n}(p)) \equiv \left|\langle \boldsymbol{n}(l_t(p))|\tilde{\rho}(p)|\boldsymbol{n}(l_b(p))\rangle_2\right|, \tag{3.11}$$

Here, $\langle\cdots\rangle_2$ is the matrix element in $2^{4S}$ dimensional Hilbert space on which the operators $\sigma_{i,\mu}^\alpha$ and $\sigma_{j,\mu}^\alpha$ operate, where $\mu = 1, 2, \cdots, 2S$ and $\alpha = x, y, z$. The operator $\tilde{\rho}(p)$ in (3.11) is defined by

$$\tilde{\rho}(p) \equiv P_{i_l(p)} P_{i_r(p)} \rho(p) P_{i_l(p)} P_{i_r(p)}. \tag{3.12}$$

with

$$\rho(p) \equiv \exp\Big(\sum_{\mu,\nu} K_p \hat{\Lambda}(\mu,\nu)\Big) \tag{3.13}$$

Here $\hat{\Lambda}(\mu,\nu)$ is defined by

$$\hat{\Lambda}_{i,j}(\mu,\nu) \equiv \sigma_{i_l(p),\mu}^x \sigma_{i_r(p),\nu}^x + \sigma_{i_l(p),\mu}^y \sigma_{i_r(p),\nu}^y + \lambda_p \sigma_{i_l(p),\mu}^z \sigma_{i_r(p),\nu}^z. \tag{3.14}$$

The constant $K_p$ in (3.13) depends on the variant of the ST-decomposition. However, in any variant of the ST-decomposition, the following equation holds:

$$\sum_{\substack{p \\ i_l(p) = i \\ i_r(p) = j}} K_p = \beta J_{i,j}/4 \qquad \text{for all } (i,j), \tag{3.15}$$

where the summation is over the plaquette whose spatial location is specified by two spatial indices $i$ and $j$.

As a function of $K_p$ and $\lambda_p$, the local weight function $w(\boldsymbol{n}(p))$ has the following property

$$[w(\boldsymbol{n}(p))]_{K_p,\lambda_p} = [w(\boldsymbol{n}(p))]_{-K_p,-\lambda_p} \tag{3.16}$$

Because of this identity, we can assume positive $K_p$ without loss of generality. With this assumption, we can rewrite (3.11) simply as

$$w(\boldsymbol{n}(p)) \equiv \langle \boldsymbol{n}(l_t(p))|\tilde{\rho}(p)|\boldsymbol{n}(l_b(p))\rangle_2 \geq 0. \tag{3.17}$$

In order to prove the identity (3.16), we notice that

$$\langle \boldsymbol{n}(l_t(p))|\tilde{\rho}(p)|\boldsymbol{n}(l_b(p))\rangle_2 = \frac{\langle\langle m_{tl}, m_{tr}|\rho(p)|m_{bl}, m_{br}\rangle\rangle}{\sqrt{\binom{2S}{m_{tl}}\binom{2S}{m_{tr}}\binom{2S}{m_{bl}}\binom{2S}{m_{br}}}}, \tag{3.18}$$

where

$$m_X \equiv \sum_{v \in s_X(p)} n_v, \qquad X = tl, tr, bl \text{ or } br, \tag{3.19}$$

and $|m_1, m_2\rangle\rangle$ is the eigenstate of $\boldsymbol{S}_i^2$, $\boldsymbol{S}_i^2$, $S_i^z$ and $S_j^z$ which satisfies

$$S_i^2|m_1, m_2\rangle\rangle = S_j^2|m_1, m_2\rangle\rangle = S(S+1)|m_1, m_2\rangle\rangle, \tag{3.20}$$

$$S_i^z|m_1, m_2\rangle\rangle = (-S + m_1)|m_1, m_2\rangle\rangle, \tag{3.21}$$

$$S_j^z|m_1, m_2\rangle\rangle = (-S + m_2)|m_1, m_2\rangle\rangle. \tag{3.22}$$

Since the term

$$S_{i_l(p)}^x S_{i_r(p)}^x + S_{i_l(p)}^y S_{i_r(p)}^y \tag{3.23}$$

in (3.13) corresponds to transferring a 'particle' (i.e. a vertex with vertex variable 1) from one side (left or right) to the opposite, a matrix element

$$\langle\langle m_{tl}, m_{tr}|\rho(p)|m_{bl}, m_{br}\rangle\rangle \tag{3.24}$$

with an odd value of $m_{tl} - m_{bl} = -m_{tr} + m_{br}$ must be an odd function of $J_x = J_y = J$. For the same reason, if $m_{tl} - m_{bl}$ is even, the matrix element must be an even function of $J$. Therefore, changing the sign of $K$ and $\lambda$ at the same time, which is equivalent to changing the sign of $J_x = J_y$ with $J_z$ fixed, may change the sign of some of matrix elements, not the absolute values. Therefore, because of (3.11) and (3.18), (3.16) follows.

In (3.10), we employed a method of decomposition in which the two-body interaction in the original Hamiltonian corresponds to a four-body interaction on the plaquettes. It is possible to decompose the partition function in such a way that the local units are cubes, for example, instead of plaquettes. This possibility will be briefly discussed in Section VI.

### B. Notation and Definitions

In Section II, we introduced several notations and definitions to present two pedagogical examples. Although most essential concepts have been presented there already, we need to extend them to describe the algorithm for more general cases in an unambiguous fashion. In this subsection, we will summarize and generalize those notations and definitions.

A *graph* $G = (V_G, E_G)$ is defined by two sets $V_G$ and $E_G$. We call elements of $V_G$ *vertices* and elements of $E_G$ *edges*. In what follows, two symbols $v$ and $e$, often with subscripts and superscripts, denote a vertex and an edge. Every edge has two attributes: two vertices at its *end-points* and *color*, i.e., *green* or *red*.

A graph $G' = (V_{G'}, E_{G'})$ is called *subgraph* of $G = (V_G, E_G)$ if $V_{G'} \subset V_G$, $E_{G'} \subset E_G$. A *path* $P$ is a special graph whose vertices and edges can be numbered in such a way that the $i$-th edge's end-points are the $i$-th and the $(i+1)$-th vertices. Two vertices $v$ and $v'$ are called *connected* in $G$ when a path in $G$ exists whose first and last vertices are $v$ and $v'$. A path is called *green* when it contains an even number of red edges. Otherwise, it is called *red*. If the first and the last vertices of a path are the same, i.e., if the path is closed, it is called a *loop*. A subgraph $C$ of $G$ is called a *cluster* if any two vertices in $C$ are connected in $G$, any edge in $G$ connecting two vertices in $C$ belongs to $C$, and no vertices or edges can be added to $C$ without violating the two previous conditions. A *union* of two graphs $G$ and $G'$ is defined by

$$G \cup G' \equiv (V_G \cup V_{G'}, E_G \cup E_{G'}). \tag{3.25}$$

In Subsection II D, we saw that a special type of edges, namely, green vertical or red horizontal ones, played a particularly important role in representing the local weight $w(\boldsymbol{n}(p))$ in the form (2.58). We will see in the next subsection that different types of edges are important for different types of anisotropy of the models. Therefore, it is useful to classify open paths and edges into several categories. They are classified according to the relative locations of their end-points: A *vertical* path connects vertices whose spatial indices are the same but have different temporal indices, a *diagonal* path connects vertices whose spatial and temporal indices are different, a *horizontal* path connects vertices whose spatial indices are different but the temporal indices are the same, and a *recurrent* path connects vertices whose spatial and temporal indices are the same. The vertical paths and the diagonal paths are both called *temporal*, whereas the horizontal paths and the recurrent paths are called *spatial*. Some of these paths are special for ferromagnetic models and some others for antiferromagnetic ones. We will call a green vertical or a green diagonal path *ferromagnetic*, and a green vertical or red horizontal path *antiferromagnetic*.

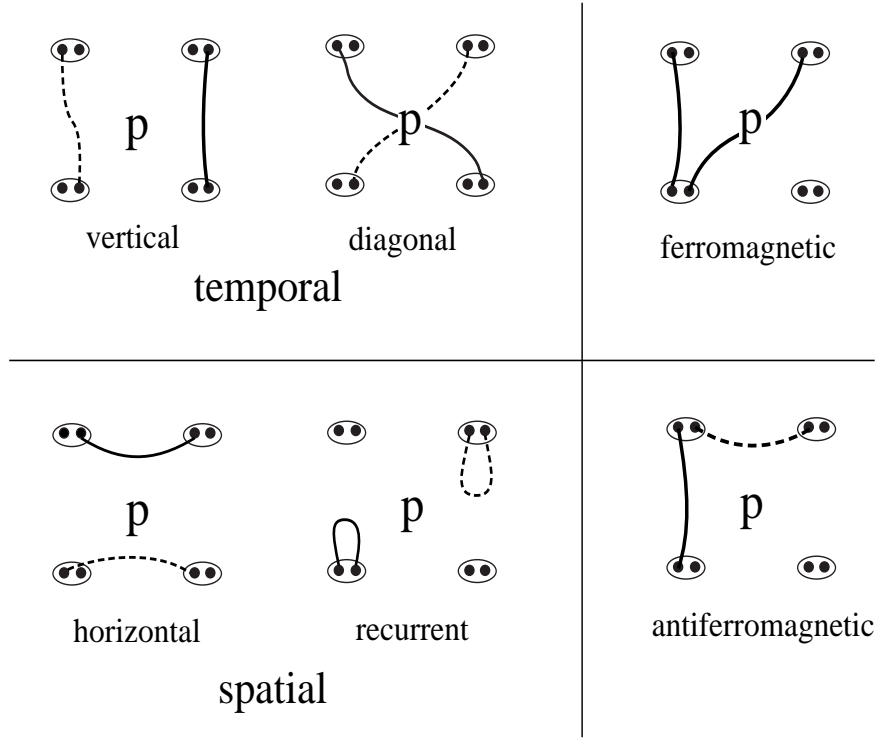

FIG. 4. The types of edges. Red edges are represented by dashed curves whereas green edges are represented by solid curves.

A green vertical path is both ferromagnetic and antiferromagnetic while a recurrent path is neither. Since a graph with only one edge and two vertices that the edge connects can be viewed as a special open path, we can apply the above classification to edges as well. These definitions are summarized in Fig. 4.

As we saw already in the last section, we define not only graphs but also a set of vertex variables $\boldsymbol{n}$ on the same vertex set as graphs. With these one-bit variables $\boldsymbol{n}$ and graphs $G$, we can express the partition function in the form of (2.1) with (2.3). Those equations can be viewed as representing a system that consists of vertex variables and graphs interacting each other. From (2.4), it is clear that the function $\Delta(\boldsymbol{n}, G)$ is the 'interaction' between vertex variables and graphs. In the two examples presented in the last section, we saw two of its special forms. Here, we restate its definition given in Subsection II D for $S = 1$ Heisenberg model in a more clear fashion and show a few of its important properties. The function $\Delta(\boldsymbol{n}(V_G), G)$ is defined by

$$\Delta(\boldsymbol{n}(V_G), G) \equiv \prod_{e \in E_G} \epsilon_e(\boldsymbol{n}(e)) \tag{3.26}$$

with a one-bit function $\epsilon_e(\boldsymbol{n}(e))$ which takes value 1 if $e$ is green and the vertex variables take the same value at the two end-points, or if $e$ is red and the variables take different values. Otherwise, the function takes value 0. If $\Delta(\boldsymbol{n}(V_G), G) = 1$, the graph $G$ and $\boldsymbol{n}(V_G)$ are called *compatible* with each other. Examples for compatible and incompatible cases are shown in Fig. 5.

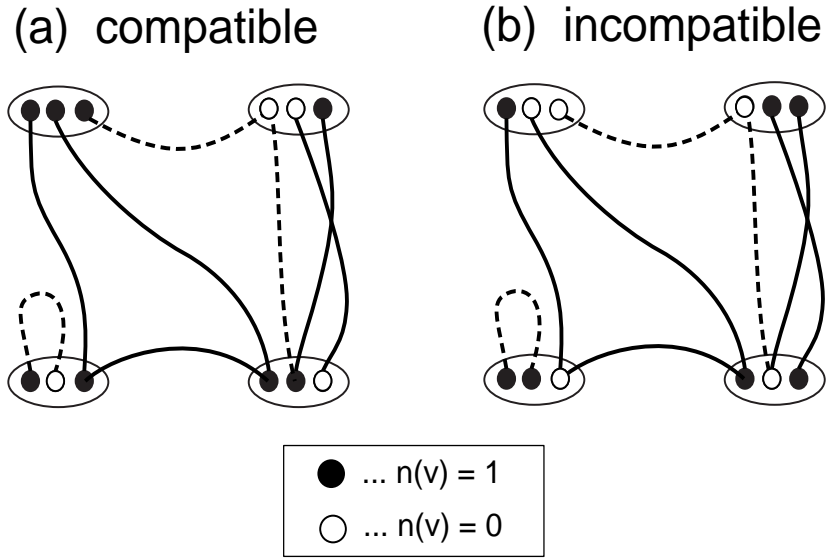

FIG. 5. Two local configurations a) compatible and b) incompatible with a given graph. The vertices for which the vertex variables are 1 are represented by solid circles, the other by open circles.

It is obvious from the definition that $\Delta(\boldsymbol{n}(V_G), G)$ is non-zero if $n_v = n_{v'}$ for any pair of vertices $v$ and $v'$ connected by a green path and if $n_v = \bar{n}_{v'} \equiv 1 - n_{v'}$ for any pair of vertices connected by a red path. For two graphs $G$ and $G'$, the following identities hold:

$$\Delta(\boldsymbol{n}(V_G), G)\Delta(\boldsymbol{n}(V_{G'}), G') = \prod_{e \in E_G} \epsilon_e(\boldsymbol{n}(e)) \prod_{e \in E_{G'}} \epsilon_e(\boldsymbol{n}(e))$$
$$= \prod_{e \in E_G \setminus E_{G'}} \epsilon_e(\boldsymbol{n}(e)) \prod_{e \in E_{G'} \setminus E_G} \epsilon_e(\boldsymbol{n}(e)) \prod_{e \in E_G \cap E_{G'}} (\epsilon_e(\boldsymbol{n}(e)))^2$$
$$= \prod_{e \in E_{G \cup G'}} \epsilon_e(\boldsymbol{n}(e)) = \Delta(\boldsymbol{n}(V_{G \cup G'}), G \cup G'). \tag{3.27}$$

In deriving this identity, we have used the fact that $\epsilon^2 = \epsilon$. This identity is useful and will be used in what follows. As a corollary, we obtain

$$\Delta(\boldsymbol{n}(V_G), G) = \prod_{C \in \mathcal{C}(G)} \Delta(\boldsymbol{n}(V_C), C). \tag{3.28}$$

where $\mathcal{C}(G)$ is the set of clusters in $G$ and $V_C$ is the vertex set of a graph $C$. The proof is straightforward using (3.27) if we note that

$$\bigcup_{C \in \mathcal{C}(G)} C = G. \tag{3.29}$$

### C. Local Weight Equation

Given the $\Delta$-functions defined in the last subsection, our task is to find a set of graphs $\Gamma(p)$ and weight of graphs $v(G(p))$ that satisfy the local weight equation

$$w(\boldsymbol{n}(p)) = \sum_{G(p) \in \Gamma(p)} v(G(p))\Delta(\boldsymbol{n}(p), G(p)), \tag{3.30}$$

for various given weights $w(\boldsymbol{n}(p))$. Once we obtain the form (3.30), by substituting it into (3.10), we can express the Boltzmann weight as

$$W(\boldsymbol{n}) = \prod_{u \in \mathcal{U}} \sum_{G(u) \in \Gamma(u)} v(G(u))\Delta(\boldsymbol{n}(u), G(u)), \tag{3.31}$$

where $v(G(l))$ for $l \in \mathcal{U}_l$ is 1 and $\Delta(\boldsymbol{n}(l), G(l))$ is $\delta_l(\boldsymbol{n}(l))$, i.e., $\Gamma(l)$ contains only one graph $G(l)$ that consists of $2S$ green vertical edges each of which connects two vertices with the same spatial indices $i$ and vertex indices $\mu$. With the use of (3.27), the above equation can be written as

$$W(\boldsymbol{n}) = \sum_{G \in \Gamma} \Big( \prod_{u \in \mathcal{U}} v(G(u)) \Big) \Big( \prod_{u \in \mathcal{U}} \Delta(\boldsymbol{n}(u), G(u)) \Big),$$
$$= \sum_{G \in \Gamma} V(G)\Delta(\boldsymbol{n}, G), \tag{3.32}$$

where $V(G) \equiv \prod_u v(G(u))$ and $G \equiv \cup_u G(u)$. Since this representation has the form discussed in Subsection II A, we can construct the cluster algorithm following the prescription given in Subsection II B.

In Subsection II D, we defined graphs $G(p)$ whose vertex set is a plaquette with 8 vertices. We also defined a special set of graphs in (2.59) over which the summation in (3.30) should be taken. Obviously, for the general $XXZ$ models, we have to define $G(p)$ on a plaquette of $8S$ vertices not on one of only 8 vertices. Furthermore, as we will see in the next section, we have to take different types of sets as $\Gamma(p)$ in (3.30). In this subsection, we present the essential result of the next section, i.e., we present explicitly the set of graphs $\Gamma(p)$ with which the local weight equation (3.30) can have meaningful solution $v(G(p))$.

In the next section, we will find that the proper set of graphs $\Gamma(p)$ depends on the anisotropy of the problem. Therefore, we will separately treat five distinct cases: 1) $XY$-like ($|\lambda_{i,j}| < 1$), 2) isotropic ferromagnetic ($\lambda_{i,j} = 1$), 3) isotropic antiferromagnetic ($\lambda_{i,j} = -1$), 4) ferromagnetic Ising-like ($\lambda_{i,j} > 1$), and 5) antiferromagnetic Ising-like ($\lambda_{i,j} < -1$) anisotropies. Corresponding to these five cases, we will define five sets of graphs. All graphs belonging to any one of these five sets share a property in common. To be more specific, all the five sets are included in a larger set of graphs

$$\Gamma^{XXZ}(p) \equiv \{\ G\ |\ V_G = p,$$
"Any cluster in $G$ have even number of vertices," and
"Any edge is green temporal or red spatial." }. (3.33)

The first condition reflects the rule that, only local configurations $\boldsymbol{n}(p)$ for which $\sum_{v \in p} n_v$ is even have non-zero weight $w(\boldsymbol{n}(p))$. In other words, as long as the first condition holds, it is guaranteed that we move from an allowed state to another allowed state by flipping a cluster. The second restriction imposed upon the edges for the $XXZ$ model reflects the fact that $z$-component of total magnetization commutes with the Hamiltonian and is conserved. Flipping those edges, green temporal or red horizontal, automatically results in a state for which this conservation law holds. As we will see in the Appendix, we do not have this restriction for the $XYZ$ model in which $z$-component is not conserved while the first condition is still valid.

The five subsets are defined as

$$\Gamma^{XY}(p) \equiv \{\ G \in \Gamma^{XXZ}(p)\ |\ \text{"No two edges in } G \text{ share a vertex."}\ \}, \tag{3.34}$$

$$\Gamma^{\text{FH}}(p) \equiv \{\ G \in \Gamma^{XY}(p)\ |\ \text{"Every edge in } G \text{ is ferromagnetic."}\ \}, \tag{3.35}$$

$$\Gamma^{\text{AFH}}(p) \equiv \{\ G \in \Gamma^{XY}(p)\ |\ \text{"Every edge in } G \text{ is antiferromagnetic."}\ \}, \tag{3.36}$$

$$\Gamma^{\text{F}}(p) \equiv \{\ G \in \Gamma^{XXZ}(p)\ |\ \text{"Every edge in } G \text{ is ferromagnetic."}\ \}, \tag{3.37}$$

$$\Gamma^{\text{AF}}(p) \equiv \{\ G \in \Gamma^{XXZ}(p)\ |\ \text{"Every edge in } G \text{ is antiferromagnetic."}\ \}. \tag{3.38}$$

We note that

$$\Gamma^{FH} = \Gamma^{XY} \cap \Gamma^{F} \tag{3.39}$$

$$\Gamma^{AFH} = \Gamma^{XY} \cap \Gamma^{AF}. \tag{3.40}$$

A typical graph for each case is illustrated for the $S = 3/2$ case in Fig. 6.

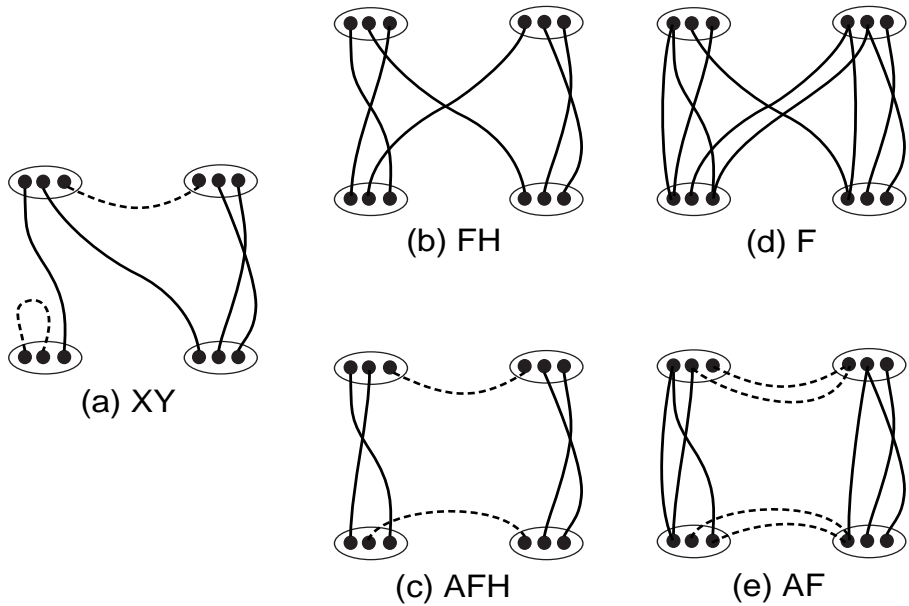

FIG. 6. Typical examples for the five classes of graphs for a) $XY$-like, b) isotropic ferromagnetic, c) isotropic antiferromagnetic, d) ferromagnetic Ising-like, and e) antiferromagnetic Ising-like anisotropies.

In the next section, we will show that with these sets of graphs, at least one non-negative set of coefficients $v(G(p))$ exists that satisfies (3.30) for each type of anisotropy. In this paper, we do not give the explicit form of those solutions except for the special case of Heisenberg models. Instead, we will present the method to compute the coefficients.

It is interesting to consider what kind of clusters will be formed when we take the union of local graphs, assuming that all plaquettes in the system have the same type of anisotropy. One striking difference is seen between the $XY$-like and the Ising-like anisotropic cases. For $XY$-like anisotropy, in a local graph $G(p)$, a vertex is an end-point of one and only one edge. This means that every vertex is shared by two and only two edges, when we take into account the fact that every site is shared by two local units, i.e., plaquettes or vertical links. Therefore, when we take the union graph $G \equiv \bigcup_u G(u)$, it consists of a number of loops without branching. Therefore, the present scheme naturally leads to a loop algorithm in the case of $XY$-like anisotropy. On the other hand, we generally have branching in the case of Ising-like anisotropy, because a vertex can be shared by any number of edges. Another important difference is that, in the case of $XY$-like and ferromagnetic Ising-like anisotropies, for any worldline in the current configuration, such a loop can form that coincides the worldline because the graphs include both vertical and diagonal edges and a worldline also consists of vertical and diagonal segments. Here a worldline is a line connecting vertices with vertex variable 1, This means that, in those cases, any worldline can vanish in a Monte Carlo step. On the other hand, in the case of antiferromagnetic Ising-like anisotropy, a worldline with diagonal segments cannot vanish because a loop which overlaps this worldline can not form. This may cause a problem concerning ergodicity as we will discuss in Section VI.

## IV. CLUSTER REPRESENTATION OF THE LOCAL BOLTZMANN FACTOR

In this section, we show that the local Boltzmann factor $w(\boldsymbol{n}(p))$, defined on a plaquette $p$, can be expressed by a sum of terms, each of which corresponds to a graph. In other words, the goal of this section is to prove the weight equation (3.30) has at least one meaningful solution with a set of graphs (3.34), (3.35), (3.36), (3.37), or (3.38) depending on the anisotropy of the problem. We also propose a method to compute the coefficients $v(G(p))$. For the general case, we will show how to compute them numerically, and for the Heisenberg models, we will provide compact formulae for the isotropic case.

Because we will focus on only one pair of interacting points in the original lattice $L$, which we refer to as $a$ and $b$, the corresponding lattice we consider in this section has just two lattice points when projected onto a horizontal plane. What we will discuss is the multiplication of two operators defined in the local Hilbert space on these two lattice points (in the original lattice). Because an operator in this space can be represented by a plaquette, it is sufficient to take two plaquettes, one stacked on the other, in order to discuss the multiplication. Therefore, the size of 'hyper-lattice' we consider here in the temporal dimension is three.

To avoid confusion, we emphasize that this hyper-lattice, whose size is two in spatial direction and three in temporal direction, is not necessarily the subset of the hyper-lattice discussed in the previous sections. The spatial indices $a$ and $b$, of course, correspond to the spatial indices in the last section, because they correspond to the lattice points in the original lattice $L$. The index $\mu$ also corresponds to the same index in the last section and specifies a vertex in a given site. However, the temporal indices in this section do not correspond to those in the last section. Here, they are merely labels introduced to distinguish one state vector from another. For example, $\boldsymbol{n}_k$ stands for $4S$ one-bit variables $n_{(k,i,\mu)}$ ($i = a, b; \mu = 1, 2, \cdots, 2S$). The state $|\boldsymbol{n}_k\rangle_2$ is an eigenvector of the $z$-components of Pauli operators, i.e.,

$$\sigma^z_{i,\mu}|\boldsymbol{n}_k\rangle_2 = (2n_{(k,i,\mu)} - 1)|\boldsymbol{n}_k\rangle_2. \quad (4.1)$$

Defining $l_k$ as a set of two sites with the temporal index $k$, i.e., $\{(k, a), (k, b)\}$, we can rewrite $\boldsymbol{n}_k$ as $\boldsymbol{n}(l_k)$. We will use this notation in what follows.

Another useful tool for the discussion is operators whose matrix elements are given by the $\Delta$-functions defined in Subsection III B. We define an operator $\hat{\Delta}(G(p))$ by

$$\langle \boldsymbol{n}(l_t(p))|\hat{\Delta}(G(p))|\boldsymbol{n}(l_b(p))\rangle_2 \equiv \Delta(\boldsymbol{n}(p), G(p)). \quad (4.2)$$

Here, $l_t(p)$ and $l_b(p)$ are the top and bottom link of the plaquette $p$ as defined in Subsection III B. The operator $\hat{\Delta}(G(p))$ depends on $p$ only through $i_r(p)$ and $i_l(p)$ and does not depend on the temporal index $k$.

To outline the proof before going into its detail is useful. We will introduce an operation called the contraction of two $\Delta$-operators in such a way that the product of two operators $\hat{\Delta}_{G_p}$ and $\hat{\Delta}_{G_p''}$ is $\hat{\Delta}_{G_p''}$ multiplied by a scalar factor where $G''$ is the graph resulting from contraction of $G_p$ and $G_p'$. In other words, the contraction of two graphs corresponds to multiplication of two operators. This relationship enables us to discuss the nature of the operators in a graph-theoretical language. The most important feature that we will prove is that a set of operators $O^*$ is closed with respect to multiplication. Here, the symbol $*$ stands for 'XY', 'FH', 'FAH', 'F' or 'AF', corresponding to XY-like, isotropic-ferromagnetic, isotropic-antiferromagnetic, ferromagnetic-Ising-like, or antiferromagnetic-Ising-like anisotropy, and $O^*$ is the set of operators $\hat{o}$ which can be expressed as a sum of elements of $\Gamma^*$ multiplied by some non-negative coefficients $v(G(p))$. Formally,

$$O^* \equiv \left\{ \hat{o} \mid \exists v(G(p)) \geq 0 \left[ \hat{o} = \sum_{G(p) \in \Gamma^*(p)} v(G(p)) \hat{\Delta}(G(p)) \right] \right\}. \tag{4.3}$$

The sets of graphs $\Gamma^*$ are the ones defined in Subsection III C. Then, we can show that $\tilde{\rho}$ in (3.12) belongs to one of these sets of operators because 1) the product of projection operator in (3.12) belongs to all the above sets of operators, 2) all the operators $\hat{\Lambda}(\mu, \nu)$ in (3.13) become elements of one of the above sets of operators when a scaler operator $x\hat{I}$ is added to it ($\hat{I}$ is the identity operator and $x$ is some real number), 3) all of the above sets of operators are closed with respect to the multiplication of two elements, the multiplication of an element by a non-negative real number, and the addition of two elements, 4) the number of graphs which correspond to distinct $\Delta$-operators is finite, and 5) the Taylor expansion series of $\tilde{\rho}$ with respect to $K_p$ converges. In short, we will show that a set of non-negative coefficients $v(G(p))$ exists that satisfies

$$\tilde{\rho} = \sum_{G(p) \in \Gamma^*} v(G(p)) \hat{\Delta}(G(p)), \tag{4.4}$$

When expressed as an equation between matrix elements, (4.4) reduces to (3.30).

### A. The XY-like anisotropy ($|\lambda_p| \leq 1$)

The first step in proving the statement (4.4) in the case where $|\lambda_p| \leq 1$ is to note the following identity:

$$\hat{\Lambda}(\mu, \nu) = \sigma^x_{a,\mu} \sigma^x_{b,\nu} + \sigma^y_{a,\mu} \sigma^y_{b,\nu} + \lambda \sigma^z_{a,\mu} \sigma^z_{b,\nu} = -1 + (1+\lambda)\hat{A}(\mu, \nu) + (1-\lambda)\hat{B}(\mu, \nu). \tag{4.5}$$

where $\hat{A}$ and $\hat{B}$ are operators defined by

$$\langle \boldsymbol{n}(l_2) | \hat{A}(\mu, \nu) | \boldsymbol{n}(l_1) \rangle_2 =$$
$$\delta(n_{(2,a,\mu)}, n_{(1,b,\nu)}) \delta(n_{(2,b,\nu)}, n_{(1,a,\mu)}) \prod_{\substack{\alpha \neq \mu \\ \beta \neq \nu}} \delta(n_{(2,a,\alpha)}, n_{(1,a,\alpha)}) \delta(n_{(2,b,\beta)}, n_{(1,b,\beta)}) \tag{4.6}$$

$$\langle \boldsymbol{n}(l_2) | \hat{B}(\mu, \nu) | \boldsymbol{n}(l_1) \rangle_2 =$$
$$\delta(\bar{n}_{(2,a,\mu)}, n_{(2,b,\nu)}) \delta(n_{(1,a,\mu)}, \bar{n}_{(1,b,\nu)}) \prod_{\substack{\alpha \neq \mu \\ \beta \neq \nu}} \delta(n_{(2,a,\alpha)}, n_{(1,a,\alpha)}) \delta(n_{(2,b,\beta)}, n_{(1,b,\beta)}). \tag{4.7}$$

Here, the symbol $\bar{n}$ denotes $1-n$. It is easy to see that the operator $\hat{A}(\mu, \nu)$ can be expressed as

$$\hat{A}(\mu,\nu) \equiv \hat{\Delta}(A(\mu,\nu)), \qquad (4.8)$$

where $A(\mu,\nu)$ is a graph shown in Fig. 7(a).

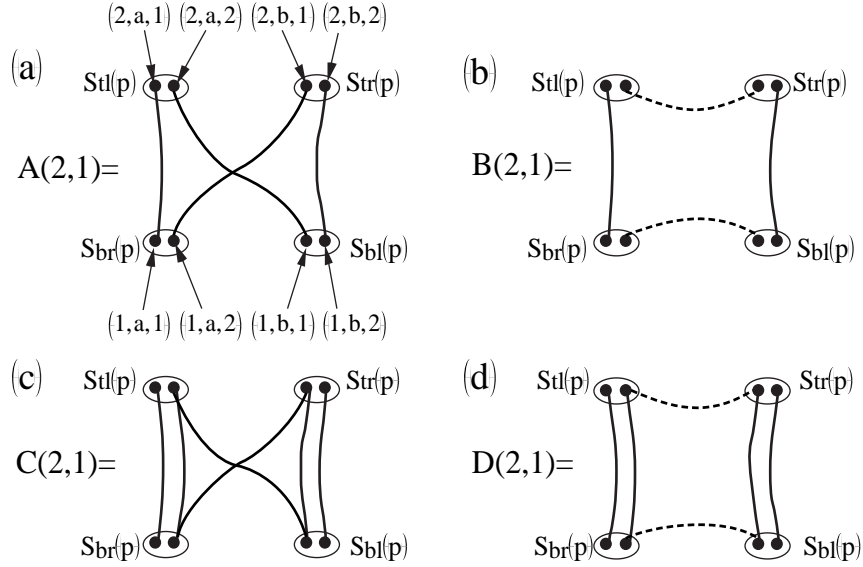

FIG. 7. Examples for the graphs which define (a) $\hat{A}$, (b) $\hat{B}$, (c) $\hat{C}$ and (d) $\hat{D}$ in the case of $S = 1$. The Hamiltonian of a $XXZ$ model can be expressed as a sum of these four types of operators. For models with $XY$-like anisotropy, only graphs of type (a) and (b) are sufficient to express the Hamiltonian whereas (b) is replaced by (c) in the case of ferromagnetic Ising-like anisotropy and by (d) in the case of antiferromagnetic Ising-like anisotropy. The ovals are sites, the solid circles are vertices and the curves are edges. In the graphs for the $XXZ$ models, all spatial edges are 'red', and all temporal edges are 'green'.

Similarly, for the operator $\hat{B}_{(\mu,\nu)}$, we have

$$\hat{B}(\mu,\nu) \equiv \hat{\Delta}(B(\mu,\nu)), \qquad (4.9)$$

with the graph $B(\mu,\nu)$ shown in Fig. 7(b).

Using the identity (4.5), we can rewrite the exponential operator in (3.13) as

$$\rho = \exp(\sum_{\mu,\nu} K_p \Lambda^{(\mu,\nu)})$$
$$= \exp\left[\frac{K_p}{4} \sum_{\mu,\nu} \left(-1 + (1+\lambda_p)\hat{A}(\mu,\nu) + (1-\lambda_p)\hat{B}(\mu,\nu)\right)\right]$$
$$= e^{-K_p S^2} \exp\left[\frac{K_p}{4} \sum_{\mu,\nu} \left((1+\lambda_p)\hat{A}(\mu,\nu) + (1-\lambda_p)\hat{B}(\mu,\nu)\right)\right]. \qquad (4.10)$$

The exponent of the last factor

$$\frac{K_p}{4} \sum_{\mu,\nu} \left((1+\lambda_p)\hat{A}(\mu,\nu) + (1-\lambda_p)\hat{B}(\mu,\nu)\right) \tag{4.11}$$

is an element of $O^{XY}$ because $K_p$, $1+\lambda_p$ and $1-\lambda_p$ are non-negative and $O^{XY}$ is obviously closed with respect to the addition of two elements and the multiplication of an element by a non-negative number. Additionally, both $A(\mu,\nu)$ and $B(\mu,\nu)$ belong to $\Gamma^{XY}(p)$, or equivalently, $\hat{A}(\mu,\nu)$ and $\hat{B}(\mu,\nu)$ belong to $O^{XY}$.

On the other hand, matrix elements of the projection operators in (3.12) can simply be written in the following form.

$$\langle \boldsymbol{n}(l_2)|P_a P_b|\boldsymbol{n}(l_1)\rangle_2 = \frac{1}{((2S)!)^2} \sum_{\pi,\pi'} \prod_{\mu=1}^{2S} \delta(n_{(2,a,\pi(\mu))}, n_{(1,a,\mu)}) \delta(n_{(2,b,\pi'(\mu))}, n_{(1,b,\mu)}). \tag{4.12}$$

where $\pi$ and $\pi'$ stand for permutations of the set $\{1, 2, \cdots, 2S\}$ and the summation is taken over all possible permutations. Therefore, this product of the projection operators can be viewed as a sum of $\Delta$-operators for which the graph consists of $2S$ green vertical edges which connect $(1, a, \mu)$ to $(2, a, \pi(\mu))$ and $(1, b, \mu)$ to $(2, b, \pi'(\mu))$. Therefore, the operator $P_a P_b$ belongs to all five sets of operators defined in (4.3). To summarize, we have shown that $\tilde{\rho}$ in (3.17) can be expressed in the form

$$\tilde{\rho} = \hat{Y} e^{\hat{X}} \hat{Y} \tag{4.13}$$

with two elements $\hat{X}$ and $\hat{Y}$ of $O^{XY}$.

We now consider an arbitrary set $\Gamma(p)$ of graphs defined on a plaquette $p$ and its two arbitrary elements, $G_1$ and $G_2$. If there exists an element $G$ in $\Gamma(p)$ for which

$$\hat{\Delta}(G_1)\hat{\Delta}(G_2) = 2^m \hat{\Delta}(G) \tag{4.14}$$

holds with some non-negative integer $m$, we call $\Gamma(p)$ *closed* with respect to multiplication. With this definition, we will prove the following statement:

**Lemma 1** $\Gamma^{XY}(p)$ *is closed with respect to multiplication.*

If we assume Lemma 1, it is obvious that $O^{XY}$ defined in (4.3) is closed with respect to multiplication.

We first assume that the above Lemma is true and show that $\tilde{\rho}$ expressed in the form (4.13) belongs to $O^{XY}$. To this end, we consider the Taylor expansion of $e^{\hat{X}}$ with respect to $\hat{X}$. We then neglect the terms of the $(n+1)$-th order and higher. The $k$-th order term in the Taylor expansion is $(1/k!)\hat{X}^k$ and this obviously belongs to $O^{XY}$. So does the $k$-th order term of $\tilde{\rho}$, i.e., $(1/k!)\hat{Y}\hat{X}^k\hat{Y}$. Thus the $n$-th order approximant of $\tilde{\rho}$ also belongs to $O^{XY}$. In other words, the operator $\tilde{\rho}$ in (3.12) can be approximated up to the $n$-th order by $\tilde{\rho}^{(n)} \in O^{XY}$. Namely,

$$\tilde{\rho} \approx \tilde{\rho}^{(n)} \equiv \sum_{G(p) \in \Gamma^{XY}(p)} v^{(n)}(G(p)) \hat{\Delta}(G(p)), \tag{4.15}$$

with non-negative coefficients $v^{(n)}(G(p))$. Here we note that the number of distinct elements of $\Gamma^{XY}(p)$ is finite. We also note that the series $v^{(n)}(G(p))$ ($n = 1, 2, 3, \cdots$) is monotonically convergent to a non-negative value because the contribution from each order term to $v^{(n)}(G)$ is non-negative. These facts and (4.15) make it obvious that in the limit of $n \to \infty$ the operator $\tilde{\rho}^{(n)}$ can still be expressed in the form (4.15), i.e., it belongs to $O^{XY}$. Thus we get the following theorem:

**Theorem 1** *If $|\lambda_p|$ is not greater than unity, the operator $\tilde{\rho}$ can be decomposed into a sum of $\Delta$-operators of the form*

$$\tilde{\rho} = \sum_{G(p) \in \Gamma^{XY}(p)} v(G(p)) \hat{\Delta}(G(p)), \tag{4.16}$$

*with non-negative coefficients $v(G(p))$.*

Taking the matrix elements of the both sides of the above equation, we have

$$w(\boldsymbol{n}(p)) = \sum_{G(p) \in \Gamma^{XY}(p)} v(G(p)) \Delta(\boldsymbol{n}(p), G(p)). \tag{4.17}$$

This is the statement that we wanted to show for $XY$-like anisotropic models.

### B. Contraction operation and proof of Lemma 1

Now we prove Lemma 1. We consider two arbitrary elements $G_1$ and $G_2$ of $\Gamma_p^{XY}$. Since $\hat{\Delta}(G_1)$ and $\hat{\Delta}(G_2)$ do not depend on a particular plaquette, as we discussed above, we assume that $G_1$ is defined on $p_1 \equiv l_1 \cup l_2$ and $G_2$ on $p_2 \equiv l_2 \cup l_3$ (see Fig. 8).

Using (3.27), we can express the matrix element of $\hat{\Delta}(G_2)\hat{\Delta}(G_1)$ as

$$\begin{aligned}
\langle \boldsymbol{n}(l_3)|\hat{\Delta}(G_2)\hat{\Delta}(G_1)|\boldsymbol{n}(l_1)\rangle_2 &= \sum_{\boldsymbol{n}(l_2)} \langle \boldsymbol{n}(l_3)|\hat{\Delta}(G_2)|\boldsymbol{n}(l_2)\rangle_2 \langle \boldsymbol{n}(l_2)|\hat{\Delta}(G_1)|\boldsymbol{n}(l_1)\rangle_2 \\
&= \sum_{\boldsymbol{n}(l_2)} \Delta(\boldsymbol{n}(p_2), G_2) \Delta(\boldsymbol{n}(p_1), G_1) \\
&= \sum_{\boldsymbol{n}(l_2)} \Delta(\boldsymbol{n}(l_3 \cup l_2 \cup l_1), G_1 \cup G_2). \tag{4.18}
\end{aligned}$$

Any cluster in the union $G_1 \cup G_2$ is a single path because a vertex in $l_1$ can only be an end-point of one and only one of edges in $G_1$, a vertex in $l_3$ can only be in $G_2$, and a vertex in $l_2$ is shared by an edge in $G_1$ and another in $G_2$. Therefore, the clusters (in this case, paths) have the following properties:

**Property 1** *The vertices on the top link ($l_3$) and on the bottom link ($l_1$), which correspond to $(3, i, \mu)$ and $(1, i, \mu)$, are end-points of paths. They cannot be intermediate points.*

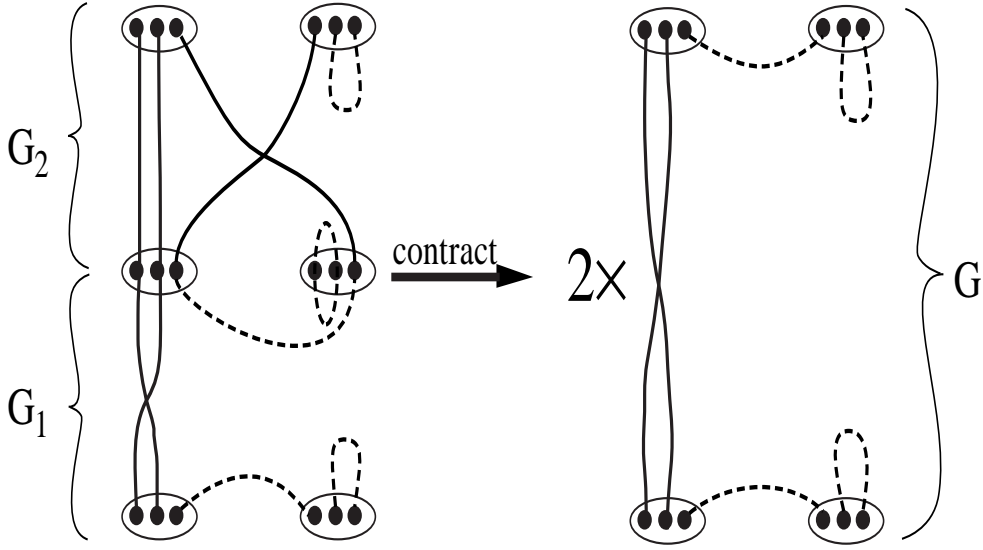

FIG. 8. Two graphs $G_1 \in \Gamma^{XY}(p_1)$ and $G_2 \in \Gamma^{XY}(p_2)$, and their contraction resulting in a graph $G$ with a factor two. The factor comes from the small loop in the right middle part of the graph on the left hand side.

**Property 2** *The vertices on the middle link ($l_2$), which corresponds to $(2, i, \mu)$, are intermediate points of paths. They cannot be end-points.*

**Property 3** *If an edge belongs to the lower graph ($G_1$), the adjacent edge(s) in the same path belongs to the upper graph ($G_2$), and vice versa.*

These elemental properties lead to the following properties:

**Property 1'** *A temporal edge can appear only as the first or the last edge in a path.*

**Property 2'** *A spatial edge can appear only as the intermediate edge in a path.*

From Property 3, the following properties are derived.

**Property 4** *The length of any loop is even.*

Here, the length of a path is the number of its edges. Similar to Property 4, for an open spatial path we have the following property:

**Property 5** *The length of any open spatial path is odd.*

This is because the first and the last edges in such an open path belong to the same graph ($G_1$ or $G_2$). For the same reason, we have

**Property 6** *The length of any temporal path is even.*

Using (3.28), the matrix elements (4.18) can be expressed in terms of contributions from these paths as

$$\langle \boldsymbol{n}(l_3)|\hat{\Delta}(G_2)\hat{\Delta}(G_1)|\boldsymbol{n}(l_1)\rangle_2 = \sum_{\boldsymbol{n}(l_2)} \prod_P \Delta(\boldsymbol{n}(V_P), P) = \prod_P f(P) \tag{4.19}$$

where $P$ stands for a path, $V_P$ is the vertex set of $P$, and $f(P)$ is the contribution from the path $P$ defined by

$$f(P) \equiv \sum_{\boldsymbol{n}(V_P \cap l_2)} \Delta(\boldsymbol{n}(V_P), P). \tag{4.20}$$

Note that the paths $P$ in (4.19) can be classified into three categories: 1) loops, 2) open spatial paths, and 3) open temporal paths. In what follows, we discuss contributions from these three types of paths separately.

We first consider the contribution in (4.19) from loops, i.e., $f(P)$ for a loop $P$. Because there are no end-points, all vertices are on the middle link $l_2$ (Property 1). Therefore, all edges are red because of the definition of $\Gamma^{XXZ}(p)$. Hence, we can write the contribution from a loop $P$ whose length is $2m$ (Property 4) in the following way:

$$f(P) = \sum_{n_1^P} \sum_{n_2^P} \cdots \sum_{n_{2m}^P} \delta(n_1^P, \bar{n}_2^P)\delta(\bar{n}_2^P, n_3^P) \cdots \delta(n_{2m-1}^P, \bar{n}_{2m}^P)\delta(\bar{n}_{2m}^P, n_1^P), \tag{4.21}$$

where $n_i^P$ is the value of the one-bit function $\boldsymbol{n}$ on the $i$-th vertex in the path $P$. The product of the $\delta$ functions is non-zero if and only if

$$n_1^P = \bar{n}_2^P = \cdots = \bar{n}_{2m}^P. \tag{4.22}$$

Therefore, a contribution from a closed path is a mere numerical factor of 2, i.e., $f(P) = 2$.

Next we consider the contribution from a spatial path. If the length of the path is one, the edge must be spatial and there is no intermediate vertex. Therefore, the contribution is $\delta(n_1^P, \bar{n}_2^P)$ and we will assume the length is larger than one. For such a path, the first and the last edges are temporal and all others are spatial, because of Properties 1 and 2. The contribution can then be written as

$$f(P) = \sum_{n_2^P} \sum_{n_3^P} \cdots \sum_{n_{2m-1}^P} \delta(n_1^P, n_2^P)\delta(n_2^P, \bar{n}_3^P)\delta(\bar{n}_3^P, n_4^P) \cdots \delta(n_{2m-2}^P, \bar{n}_{2m-1}^P)\delta(\bar{n}_{2m-1}^P, \bar{n}_{2m}^P)$$

$$= \delta(n_1^P, \bar{n}_{2m}^P). \tag{4.23}$$

Therefore, by tracing out the intermediate variables, this path is transformed into a red spatial edge in the resulting graph, whether the length is one or larger.

Finally we consider the contribution from a temporal path. In this case, the first and the last edges are temporal as the previous case. All other intermediate edges are spatial. Since the length is even (Property 6) in this case, the contribution is

$$f(P) = \sum_{n_2^P} \sum_{n_3^P} \cdots \sum_{n_{2m}^P} \delta(n_1^P, n_2^P)\delta(n_2^P, \bar{n}_3^P)\delta(\bar{n}_3^P, n_4^P) \cdots \delta(\bar{n}_{2m-1}^P, n_{2m}^P)\delta(n_{2m}^P, n_{2m+1}^P)$$

$$= \delta(n_1^P, n_{2m+1}^P). \tag{4.24}$$

Hence the contraction results in a green vertical or green diagonal edge, i.e., a green temporal edge. In Fig. 8, we can see examples of the three types of contribution we have discussed.

Thus we have shown that the right hand side of (4.19) is a product of three kinds of factors: 1) the numerical factor $2^m$ where $m$ is the number of loops, 2) the factor which corresponds to antiferromagnetic edges, and 3) the factor which corresponds to ferromagnetic edges. Hence Lemma 1 is proven.

In this subsection, we traced out the intermediate variables to obtain a new $\Delta$-function with a multiplicative factor. As illustrated in Fig. 8, it can be done graphically by 1) erasing intermediate vertices, 2) replacing open paths by edges with the same color, and 3) replacing each loop by a factor 2. In what follows, we call this graphical operation a *contraction*.

### C. A numerical method to compute the coefficients $v(G_p)$

The discussion based on the Taylor expansion that led (4.15) provides us not only a proof of Theorem 1 but also a numerical method for computing $v(G) \equiv \lim_{n \to \infty} v^{(n)}(G)$. What we have to do is

1. Make a table of the multiplication rules among the elements of $\Gamma^{XY}(p)$, i.e., a table which tells us which graph and what scaler multiplicative factor result from contraction of two arbitrary graphs in $\Gamma^{XY}(p)$.

2. Compute the Taylor expansion of the exponential operator in (4.10) up to the $n$-th order using the table.

3. Compute the result of multiplication of the projection operators. Or, equivalently, symmetrize the expression obtained in the last procedure with respect to the vertex indices.

Considering the fact that expansion series of an exponential function of bounded matrices such as (4.10) generally converges quickly, $v^{(n)}(G(p))$ in (4.15) should be good approximants for $v(G(p))$ with not too large $n$. We remark that the above procedure does not become combinatorially difficult as $n$ increases. The amount of computation is only proportional to $n$. Therefore, this procedure to compute the coefficient is reasonably practical, although it may not be optimal. We also remark that the above procedure applies to other types of anisotropy to be now discussed.

### D. The isotropic cases ($|\lambda_p| = 1$)

We now consider the special cases where $\lambda_p = 1$ or $\lambda_p = -1$. These cases have already been considered in the last subsection. However, we take up these cases separately because the set of graphs we must consider for the expression (4.16) is truly smaller than $\Gamma^{XY}(p)$. This reduction means that the set of graphs needed in the labeling process of the simulation can be simpler than in the general case. It will also be clear in the next section that in these special cases the coefficient can even be computed analytically without resorting to the numerical method discussed above. For the isotropic cases, the sets of graphs $\Gamma^{\text{FH}}(p)$ and $\Gamma^{\text{AFH}}(p)$ play the same role as $\Gamma^{XY}(p)$ did for the $XY$-like anisotropy. In what follows, we call a graph which belongs to $\Gamma^{\mathrm{FH}}(p)$ a *special ferromagnetic* graph, and one which belongs to $\Gamma^{\mathrm{AFH}}(p)$, a *special antiferromagnetic* graph.

We can show for the special ferromagnetic graphs that

**Lemma 2** $\Gamma^{\mathrm{FH}}(p)$ *is closed with respect to multiplication.*

It is sufficient to prove that the following property exists in the union graph defined on $p_1 \cup p_2$:

**Property 7** *Given a graph defined on two plaquettes sharing a link, if all edges in a path are ferromagnetic, the path is also ferromagnetic.*

As we did in the last subsection, we consider two plaquettes stacked on top of each other. In the present case, we consider a special ferromagnetic graph defined on each plaquette. To show Property 7, it is sufficient to prove that we can obtain only temporal paths from ferromagnetic edges and that those paths are green. Because of Property 1', any possible path in the stacked special ferromagnetic graphs consists of two temporal edges: one in the lower graph and the other in the upper graph. Therefore, the resulting path must be temporal. Because these two edges are both green, the resulting path is also green. Thus, Lemma 2 is proven. As a corollary, we can easily show that $O^{\mathrm{FH}}$ is closed with respect to multiplication.

A statement similar to Lemma 2 holds for the antiferromagnetic graphs:

**Lemma 3** $\Gamma^{\mathrm{AFH}}(p)$ *is closed with respect to multiplication.*

This is equivalent to the following statement.

**Property 8** *Given a graph defined on two plaquettes sharing a link, if all edges in a path are antiferromagnetic, the path is also antiferromagnetic.*

It is sufficient to prove that there are no diagonal or recurrent paths in the stacked special antiferromagnetic graphs and that all horizontal paths are red while all vertical paths are green. First, we assume that a path $P$ is diagonal. Because all edges are either horizontal or vertical, we must then have an odd number of horizontal edges and two vertical edges in $P$. Therefore, the length of $P$ must be odd. However, this contradicts Property 6 so there are no diagonal paths. Next, we assume that a path $P$ is recurrent. With the same reason as above, the length of $P$ must be even. However, this contradicts Property 5 so no recurrent edge can appear. As for the colors of the resulting path, horizontal ones must be red while vertical ones must be green because of Lemma 1 and the fact that $\Gamma_p^{\mathrm{AFH}} \subset \Gamma_p^{XY}$. Thus, Lemma 3 is proven. It follows that $O^{\mathrm{AFH}}$ is closed with respect to multiplication.

Because the graphs for the projection operators (4.12) belong to $\Gamma^{\mathrm{FH}}(p)$, by using Lemma 2 and following the same line of the argument that proved Theorem 1, we can expand the operator $\tilde{\rho}$ in terms of ferromagnetic $\Delta$-operators:

**Theorem 2** *The ferromagnetic operator $\tilde{\rho}$ can be decomposed into a sum of $\Delta$-operators of the form*

$$\tilde{\rho} = \sum_{G(p) \in \Gamma^{\mathrm{FH}}(p)} v(G(p)) \hat{\Delta}(G(p)), \qquad (4.25)$$

*with non-negative coefficients $v(G(p))$.*

By the same argument, using Lemma 3, we have the following theorem for the antiferromagnetic operator $\tilde{\rho}$:

**Theorem 3** *The antiferromagnetic operator $\tilde{\rho}$ can be decomposed into a sum of $\Delta$-operator of the form*

$$\tilde{\rho} = \sum_{G(p) \in \Gamma^{\text{AFH}}(p)} v(G(p))\hat{\Delta}(G(p)), \tag{4.26}$$

*with non-negative coefficients $v(G(p))$.*

### E. The ferromagnetic Ising-like anisotropy ($\lambda \geq 1$)

In the case of Ising-like anisotropy, we cannot use expression (4.5) as a starting point, because the coefficient $(1 - \lambda)$ may be negative. Therefore, instead of (4.5), we consider the following identity with positive coefficients as the new starting point for the ferromagnetic case ($\lambda \geq 1$):

$$\sigma^x_{a,\mu}\sigma^x_{b,\nu} + \sigma^y_{a,\mu}\sigma^y_{b,\nu} + \lambda\sigma^z_{a,\mu}\sigma^z_{b,\nu} = -\lambda + 2\hat{A}(\mu,\nu) + 2(\lambda-1)\hat{C}(\mu,\nu), \tag{4.27}$$

where the new operator $\hat{C}(\mu,\nu)$ is defined by

$$\langle \bm{n}(l_2)|\hat{C}(\mu,\nu)|\bm{n}(l_1)\rangle_2 =$$
$$\delta(n_{(2,a,\mu)}, n_{(1,a,\mu)})\delta(n_{(2,b,\nu)}, n_{(1,b,\nu)})\delta(n_{(2,a,\mu)}, n_{(1,b,\nu)})\delta(n_{(2,b,\nu)}, n_{(1,a,\mu)})$$
$$\times \prod_{\alpha \neq \mu} \delta(n_{(2,a,\alpha)}, n_{(1,a,\alpha)}) \prod_{\beta \neq \nu} \delta(n_{(2,b,\beta)}, n_{(1,b,\beta)}). \tag{4.28}$$

The graph $C(\mu,\nu)$ which expresses this operator is shown in Fig. 7(c). With this graph, we can simply write $\hat{C}(\mu,\nu) \equiv \hat{\Delta}(C(\mu,\nu))$. Obviously, the new graph does not correspond to any graph in $\Gamma^{XY}(p)$ because some of vertices are shared by multiple edges. However, it still belongs to $\Gamma^{XXZ}(p)$. In fact, $A(\mu,\nu)$ and $C(\mu,\nu)$ both belong to $\Gamma^F(p)$ defined in (3.37). In what follows, we call such a graph *ferromagnetic*. We also call the corresponding $\Delta$-operator ferromagnetic. The difference from $\Gamma^{\text{FH}}_p$ is that for $\Gamma^F_p$ there is no restriction on the number of edges sharing a vertex.

For the ferromagnetic graphs, we can show the following statement:

**Lemma 4** $\Gamma^F(p)$ *is closed with respect to multiplication.*

We will consider the contraction of these operators. We first take two ferromagnetic graphs stacked on top of one another. This time a cluster is not necessarily a path. In addition, not all the paths satisfy the Property 3. We will call paths that satisfy the Properties 1, 2 and 3 *alternating* paths. For alternating path all the properties (Properties 1-8) holds since Properties 4-8 are derived from Properties 1-3. In particular, Property 7 holds also for these alternating paths. It is important to notice that any ferromagnetic graph is equal to the union of all the alternating paths in it. Since the alternating paths overlap each other in the present case, additional consideration is needed to find the contribution from the paths when they are contracted. The result is, however, the same: a ferromagnetic alternating path contracts to a ferromagnetic edge.

To see this, we resort to an example rather than rigorous argument. The example is shown in Fig. 9.

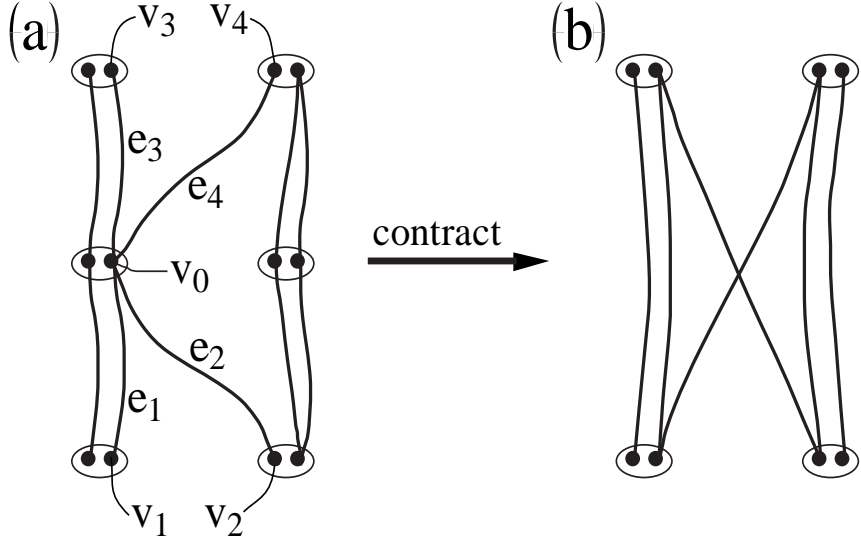

FIG. 9. An example of contraction of two ferromagnetic graphs.

In Fig. 9(a), a typical graph is shown in the case of $S = 1$. There are three clusters in the graph. The result of the contraction is merely the product of corresponding three factors. Since the left and the right clusters' contributions are trivial, we focus on the middle cluster whose vertex set is $\{v_0, v_1, v_2, v_3, v_4\}$. There are four alternating paths in this cluster. Their edge sets are $\{e_1, e_3\}$, $\{e_1, e_4\}$, $\{e_2, e_3\}$ and $\{e_2, e_4\}$. Correspondingly, the contraction of this cluster results in

$$\sum_{n_0} \delta(n_0, n_1)\delta(n_0, n_2)\delta(n_0, n_3)\delta(n_0, n_4)$$
$$= \sum_{n_0} [\delta(n_1, n_0)\delta(n_0, n_3)][\delta(n_1, n_0)\delta(n_0, n_4)][\delta(n_2, n_0)\delta(n_0, n_3)][\delta(n_2, n_0)\delta(n_0, n_4)]$$
$$= \delta(n_1, n_3)\delta(n_1, n_4)\delta(n_2, n_3)\delta(n_2, n_4), \qquad (4.29)$$

where $n_i \equiv n_{v_i}$. The last expression in the above equation is represented by Fig. 9(b). It is easy to extract essential points from this example and construct a rigorous proof. Thus we obtain Lemma 4.

Combining Lemma 4 and the expression (4.27), we obtain the following theorem:

**Theorem 4** *For $\lambda$ greater than or equal to 1, we can expand the operator $\tilde{\rho}$ in the form*

$$\tilde{\rho} = \sum_{G(p) \in \Gamma^{\mathrm{F}}(p)} v(G(p))\hat{\Delta}(G(p)), \qquad (4.30)$$

*with non-negative coefficients $v(G(p))$.*

We can compute the coefficients $v(G(p))$ numerically in this case, too, by the procedure described in Subsection IV C. The first step may seem impossible to do in the present case where we have infinitely many distinct graphs. However, even if the number of graphs is infinite, it is still possible if the number of graphs which correspond to distinct $\Delta$-operators is finite. In such a case, by identifying all the distinct graphs which give the same operator, we can keep the first step of the procedure manageable. In fact, the number of distinct $\Delta$-operators is finite in the present case. Therefore, the procedure discussed in Subsection IV C can apply.

### F. The antiferromagnetic Ising-like anisotropy ($\lambda \leq -1$)

In this case, we start with the following identity.

$$\sigma^x_{a,\mu}\sigma^x_{b,\nu} + \sigma^y_{a,\mu}\sigma^y_{b,\nu} + \lambda\sigma^z_{a,\mu}\sigma^z_{b,\nu} = \lambda + 2\hat{B}(\mu,\nu) + 2(-\lambda-1)\hat{D}(\mu,\nu), \tag{4.31}$$

where the new operator $\hat{D}(\mu,\nu)$ is defined by

$$\begin{aligned}
\langle \boldsymbol{n}(l_2)|\hat{D}(\mu,\nu)|\boldsymbol{n}(l_1)\rangle_2 = \\
\delta(n_{(2,a,\mu)}, n_{(1,a,\mu)})\delta(n_{(2,b,\nu)}, n_{(1,b,\nu)})\delta(n_{(1,a,\mu)}, \bar{n}_{(1,b,\nu)})\delta(n_{(2,b,\nu)}, \bar{n}_{(2,a,\mu)}) \\
\times \prod_{\alpha \neq \mu}\delta(n_{(2,a,\alpha)}, n_{(1,a,\alpha)}) \prod_{\beta \neq \nu}\delta(n_{(2,b,\beta)}, n_{(1,b,\beta)}).
\end{aligned} \tag{4.32}$$

The graph which expresses this operator is shown in Fig. 7(d).

In the last subsection, we defined $\Gamma^{\mathrm{F}}(p)$ by removing the restriction on the multiple occupation of vertices from $\Gamma^{\mathrm{FH}}(p)$. Similarly, in the present case, we consider a graph which consists of an arbitrary number of antiferromagnetic edges. Every vertex must be an end-point of one or more edges. We call this type of graph *antiferromagnetic* and represent the set of the antiferromagnetic graphs by $\Gamma^{\mathrm{AF}}(p)$. For the antiferromagnetic operators, we have the following lemma:

**Lemma 5** $\Gamma^{\mathrm{AF}}_p$ *is closed with respect to multiplication.*

The proof can be done in the same way as the one in the last subsection. In other words, using Property 8, we can show that all the alternating paths in the graph $G_1 \cup G_2$ are antiferromagnetic. The contribution from each alternating path can be represented by an antiferromagnetic edge, unless the path is a loop. If the path is a loop which is connected to an open path, it contributes a factor 1. If the path is a loop which is not connected to any open path, it contributes a factor 2 together with all other loops connected to it. Therefore, contraction results in antiferromagnetic edges and numerical factors. Thus follows the lemma.

Combining Lemma 5 and the expression (4.31), we obtain the following theorem:

**Theorem 5** *For $\lambda$ smaller than or equal to $-1$, we can expand the operator $\tilde{\rho}$ in the form*

$$\tilde{\rho} = \sum_{G(p) \in \Gamma^{\mathrm{AF}}(p)} v(G(p))\hat{\Delta}(G(p)), \tag{4.33}$$

*with non-negative coefficients $v(G(p))$.*

## V. HEISENBERG MODELS

In the isotropic cases, i.e., $|\lambda_p| = 1$, we can obtain compact formulae for the coefficients $v(G(p))$. To obtain this formula, we have to examine the nature of the coefficients in more detail. As we did in Subsection II D, we can classify local configurations into several categories by using symmetry properties. The function $w(\boldsymbol{n}(p))$ has various symmetries: invariance under 1) the vertical and horizontal mirror-image transformation, 2) the simultaneous flipping of all spins, and 3) the permutation of vertices within each site. We define classes of configurations $\boldsymbol{n}(p)$ in such a way that two configurations $\boldsymbol{n}(p)$ and $\boldsymbol{n}'(p)$ belong to the same class if and only if they can be transformed into each other by these symmetry transformations. We use symbol $\mathcal{S}(p)$ to specify such a class, which is a special set of states. For the same reason, we expect that the solution $v(G(p))$ possesses the same symmetry, i.e., $v(G'(p)) = v(G(p))$ if a graph $G'(p)$ can be obtained from $G(p)$ through these symmetry transformations. Therefore, it is convenient to define classes of graphs. We say that $G'(p)$ belongs to the same class as $G(p)$ if and only if $G(p)$ can be transformed into $G'(p)$ through the symmetry transformations. A class is a subset of $\Gamma^{XXZ}(p)$ and denoted by the symbol $\mathcal{G}(p)$. Note that

$$\bigcup_{\mathcal{G}(p) \subset \Gamma^*} \mathcal{G}(p) = \Gamma^*(p), \tag{5.1}$$

where $*$ stands for 'XY', 'FH', 'AFH','F' or 'AF'. Taking these definitions into account, we can reduce (3.30) to

$$\tilde{w}(\mathcal{S}(p)) = \sum_{\mathcal{G} \subset \Gamma^*(p)} N(\mathcal{S}(p), \mathcal{G}(p)) \tilde{v}(\mathcal{G}(p)), \tag{5.2}$$

where

$$\tilde{w}(\mathcal{S}(p)) \equiv w(\boldsymbol{n}(p)), \qquad \boldsymbol{n}(p) \in \mathcal{S}(p), \tag{5.3}$$
$$\tilde{v}(\mathcal{G}(p)) \equiv v(G(p)), \qquad G(p) \in \mathcal{G}(p), \tag{5.4}$$

and

$$N(\mathcal{S}(p), \mathcal{G}(p)) \equiv \sum_{G(p) \in \mathcal{G}(p)} \Delta(\boldsymbol{n}(p), G(p)), \qquad \boldsymbol{n}(p) \in \mathcal{S}(p). \tag{5.5}$$

We first note that the class of graphs $\mathcal{G}(p)$ is characterized by 16 integers, $m_{X_1-X_2}$ ($X_1, X_2 = bl, br, tl$, and $tr$). The integer $m_{X_1-X_2}$ is the number of edges which connect vertices in the site $s_{X_1}(p)$ to vertices in the site $s_{X_2}(p)$. For example, $m_{bl-tr}$ is the number of vertical edges which connects a vertex in the bottom-left site and a vertex in the top-right site. In the case of ferromagnetic Heisenberg model, these integers vanish except for $m_{bl-tl}$, $m_{bl-tr}$, $m_{br-tl}$ and $m_{br-tr}$. Since there are constraints

$$m_{bl-tl} + m_{bl-tr} = m_{br-tl} + m_{br-tr} = m_{bl-tl} + m_{br-tl} = m_{bl-tr} + m_{br-tr} = 2S, \tag{5.6}$$

only one of the above four numbers is independent. Therefore, we take $m_{bl-tl}$ as the representative. Thus, we can simply express the coefficient $\tilde{v}(\mathcal{G}(p))$ as

$$\tilde{v}(\mathcal{G}(p)) = \tilde{v}(m_{bl-tl}). \tag{5.7}$$

Next, we notice that a class of configurations is characterized by four integers, $m_{bl}, m_{br}, m_{tl}$ and $m_{tr}$, where $m_{bl}$ is the sum of vertex variables at the bottom-left site, and the other three are similarly defined. This time, three of them are independent. We express the local weight function as in Subsection II D:

$$\tilde{w}(\mathcal{S}(p)) = \tilde{w}\begin{pmatrix} m_{\rm tl} & m_{\rm tr} \\ m_{\rm bl} & m_{\rm br} \end{pmatrix} \tag{5.8}$$

Using this notation, we can rewrite (4.25) as

$$\tilde{w}\begin{pmatrix} m_{\rm tl} & m_{\rm tr} \\ m_{\rm bl} & m_{\rm br} \end{pmatrix} = \sum_{m_{\rm bl-tl}} N_{\rm FH}\begin{pmatrix} m_{\rm tl} & m_{\rm tr} \\ m_{\rm bl} & m_{\rm br} \end{pmatrix} m_{\rm bl-tl} \tilde{v}(m_{\rm bl-tl}) \tag{5.9}$$

We now consider a configuration which belongs to a class characterized by $(m_{\rm tl}, m_{\rm tr}, m_{\rm bl}, m_{\rm br}) = (m, 2S - m, 0, 2S)$. If a graph $G$ is compatible with this configuration, all $m$ occupied vertices in the top-left site must be connected to vertices in the bottom-right site because all edges in the graph are temporal and green. For a similar reason, all $2S - m$ occupied vertices in the top-right site must be connected to those in the bottom-right site. Therefore, we can easily see that the function $N$ defined above must have the following property:

$$N_{\rm FH}\begin{pmatrix} m & 2S-m \\ 0 & 2S \end{pmatrix} m_{bl-tl} = \begin{cases} ((2S)!)^2, & \text{if } m_{bl-tl} = m, \\ 0, & \text{otherwise}. \end{cases} \tag{5.10}$$

Having this equation, we can derive from (5.9) the following equation which determines $v_{\rm FH}$ completely.

$$\tilde{w}\begin{pmatrix} m & 2S-m \\ 0 & 2S \end{pmatrix} = ((2S)!)^2 \tilde{v}(m). \tag{5.11}$$

Using the identity (3.18), we finally get

$$\tilde{v}(m) = \frac{1}{((2S)!)^2}\binom{2S}{m}^{-1} \langle\langle m, 2S-m | \exp(-\Delta\tau\mathcal{H}_{i,j}) | 0, 2S\rangle\rangle. \tag{5.12}$$

Therefore, the problem has been reduced to the diagonalization of $\exp(-\Delta\tau\mathcal{H}_{i,j})$, a matrix in $(2S + 1)$ dimensions, which is numerically simple and also can be done rigorously by various symbolic computational languages.

Now, we will discuss the isotropic antiferromagnetic case. In this case, too, the class of graphs $\mathcal{G}(p)$ is characterized by four integers, $m_{bl-tl}$, $m_{br-tr}$, $m_{bl-br}$ and $m_{tl-tr}$. As in the case of ferromagnetic Heisenberg model, only one of the above four numbers are independent. We express the coefficient $\tilde{v}(\mathcal{G}(p))$ as

$$\tilde{v}(\mathcal{G}(p)) = \tilde{v}(m_{bl-tl}) \tag{5.13}$$

Equation (4.26) is rewritten as

$$\tilde{w}\begin{pmatrix} m_{tl} & m_{tr} \\ m_{bl} & m_{br} \end{pmatrix} = \sum_{m_{bl-tl}} N_{\text{AFH}}\begin{pmatrix} m_{tl} & m_{tr} \\ m_{bl} & m_{br} \end{pmatrix} m_{bl-tl} \tilde{v}(m_{bl-tl}), \qquad (5.14)$$

We next consider a configuration which belong to a class characterized by $(m_{tl}, m_{tr}, m_{bl}, m_{br}) = (0, m, m, 0)$. In this case, in any graph compatible to this configuration, all the $m$ occupied vertices in the bottom-left site must be connected to vertices in the bottom-right site because we can not connect them by green edges to the top-left site where there are no occupied vertices. Hence, we obtain the following property of $N_{\text{AFH}}$:

$$N_{\text{AFH}}\begin{pmatrix} 0 & m \\ m & 0 \end{pmatrix} m_{bl-tl} = \begin{cases} ((2S)!)^2 & \text{if } m_{bl-tl} = 2S - m, \\ 0 & \text{otherwise.} \end{cases} \qquad (5.15)$$

Accordingly, we obtain

$$\tilde{v}(m) = \frac{1}{((2S)!)^2}\binom{2S}{m}^{-1} \langle\langle 0, m | \exp(-\Delta\tau \mathcal{H}_{i,j}) | m, 0 \rangle\rangle. \qquad (5.16)$$

This is the generalization of (2.62).

## VI. ERGODICITY

Here we call an algorithm ergodic if an arbitrary state can be reached with a non-zero probability within a finite number of Monte Carlo steps regardless of the initial state. The ergodicity and the detailed balance condition are two important features that any algorithm must possess. In the main text of this paper, we focused on the detailed balance condition and left out the discussion of ergodicity. It is sometimes non-trivial to show that an algorithm possesses this property. For example, the ergodicity of conventional algorithms has not yet been proved as far as we know. However, for the new algorithm presented in this paper, it is almost straightforward to prove that ergodicity holds in the case of the ferromagnetic and $XY$-like models, as we will see below. In the antiferromagnetic models, we can show that the new algorithm is ergodic if the conventional algorithm is.

The main part of conventional worldline algorithms is local loop flips where size and shape of the loops are fixed. In order to achieve the ergodicity, the shape and the location must be chosen carefully. Even if we choose them properly, however, using only the local movements does not constitute an ergodic algorithm because some quantities are conserved by those movements in the conventional algorithm [20]. Global winding numbers are such conserved quantities. Therefore, we have to include several different kinds of global updates that can change the winding numbers. What local updates and global updates exactly we should include depends on the model. We always have to face this annoying question as long as we use the conventional algorithms. In fact, this difficulty also makes the actual computer programs complicated in order to accommodate different kinds of procedures.

Now we prove the ergodicity of the new algorithm in the ferromagnetic and $XY$-like cases. To this end, we consider an arbitrary worldline configuration. If we note that any worldline consists of vertical line segments and diagonal ones, it is easy to realize that in a labeling process a loop configuration can be generated with a finite probability in such a way that any worldline in the initial state coincides one of the loops in the loop configuration.

(Of course this is possible only in the case of ferromagnetic or $XY$-like systems where graphs with diagonal edges can be chosen.) Once such loops are formed, in the subsequent flipping process, it can happen that all existing worldlines vanish (flip) and no new worldlines are created. The outcome is the vacuum state. Thus, within one Monte Carlo step, the vacuum state can be reached from an arbitrary state. The inverse process can also take place with a finite probability, i.e., the transition from the vacuum state to an arbitrary state. Therefore, we can conclude that every state can be reached from any state within two Monte Carlo steps via the vacuum state. Thus, the ergodicity is proven.

This proof was possible because vertical and diagonal segments appear in the graphs corresponding to the operators $\hat{I}$ and $\hat{A}$ in the decomposition (4.5) and (4.27). Since the decomposition (4.31) does not contain the operator $\hat{A}$, the ergodicity of the antiferromagnetic model is not obvious. However, we can argue that if conventional algorithms are ergodic, the new algorithm is also ergodic. We can see this simpy by noticing that most global flips introduced in the conventional algorithms can happen in the new algorithm with finite probability. To be more specific, the $n$-direction global flips in [20] are simply flipping of loops with the temporal winding number 1, and the $x$-direction global flips are those of loops with spatial winding number 1. Both types of loops can form in the new algorithm. The $t$-direction global flips are equivalent to flipping of loops whose temporal and spatial winding numbers are 1, as far as the effect on the global winding numbers is concerned. Again, this type of loops can form in the new algorithm.

Here, we emphasize that these global movements are intrinsically included in the new algorithm and they do not introduce any additional complication into the algorithm. In other words, we can achieve the ergodicity in a simple and systematic way in the new algorithm in contrast to the conventional algorithms.

## VII. SUMMARY

In this paper, we described how the Boltzmann factor of the lattice quantum spin model can be decomposed into sum of terms each of which corresponds to a graph. Based on this decomposition, we have shown that a non-trivial cluster algorithm exists for any quantum spin model which can be described by this Hamiltonian, regardless of geometrical properties of the original lattice such as the number of dimensions, the boundary condition, and of the range of the interactions. This decomposition also determines how to choose the proper set of local graphs and how to assign one of them to each plaquette. The new algorithm is advantageous for several reasons: 1) it may reduce the autocorrelation time drastically, 2) with the improved estimators [19], it can reduce variances of distribution functions of important physical quantities and therefore reduce the statistical error, 3) it can achieve ergodicity without introducing any ad hoc global updates, and 4) the resulting computer programs can be simpler than those for the conventional algorithms. For some cases, such as the Ising model and the $S = 1$ antiferromagnetic, we explicitly gave the labeling probabilities that determine the Monte Carlo algorithm. For the general case, a method for computing the labeling probabilities numerically was presented.

Our representation can be viewed as the extension of the FK cluster representation of the Ising model. Therefore, the resulting Monte Carlo algorithms are generalizations of the SW algorithm to the quantum spin problems. As we showed in [7], the present scheme converges to SW algorithm in the limit of strong Ising-like anisotropy. The present algorithm also includes the loop algorithm for the $S = 1/2$ model [15].

It is remarkable that in contrast to the conventional algorithm the cluster algorithm must be quite different depending on the anisotropy of the model. For the $XY$-like anisotropy, we can have a loop algorithm. A loop algorithm is advantageous because we can identify loops in a computational time proportional to the number of vertices $N_v = 2SM|L|$. For the Ising-like anisotropy, we inevitably create clusters with branching. In this case, the computational time is proportional to $N_v \log(N_v)$.

Although the efficiency of the algorithms has not yet been demonstrated systematically, some encouraging facts are already known. The efficiency of the SW algorithm near a critical points is well-known. Other examples are the efficiency of the algorithms for the $S = 1/2$ antiferromagnetic Heisenberg model [16] in two dimensions and $S = 1$ antiferromagnetic Heisenberg model in one dimension [7]. In the latter, the autocorrelation time was at least three orders of magnitude smaller than that of the conventional method. The difference tends to be greater for large Trotter numbers. We also found that the present algorithm dramatically reduces the autocorrelation time of $S = 1/2$ $XY$ model in two dimensions, which is equivalent to hard-core boson model. This result will be published elsewhere [21]. Based on these findings, we believe that for the homogeneous $XXZ$ spin model without a symmetry breaking field such as a magnetic field, the present algorithm provides a more efficient alternative to the standard local Metropolis algorithm. On the other hand, we know much less about the efficiency of the algorithm for disordered system and systems with symmetry breaking fields. As for the disorder, it is known [3] that the SW algorithm is not advantageous for random-bond Ising model, i.e., the Edwards-Anderson spin glass model. As for the effect of the symmetry breaking field, we empirically found in the case of the $S = 1/2$ models in two dimensions that the algorithm works well for the Ising model regardless of the strength of a magnetic field whereas it does not for the $XY$ model when the field in the $z$-direction is strong. More work is needed for these cases.

## ACKNOWLEDGMENTS

The authors are grateful to G. Baker, Jr. for his critical reading of the manuscript and many suggestions. One of the authors (N.K.) thanks G. McNamara and B. Luce for useful comments on the proofs of the theorems. The work of J.E.G. was supported by the High Performance Computing and Communication program of the Department of Energy.## APPENDIX: THE COMPLETELY ANISOTROPIC CASE — THE XYZ MODEL

Since the case where $J_x = J_y$ is much more frequently studied than completely asymmetric case where $J_x \neq J_y \neq J_z \neq J_x$, we focused on the $XXZ$ model in the main text to avoid too much complication. However, the $XYZ$ model can be treated in a fashion similar to that presented for the $XXZ$ model. In this appendix, we give a brief outline of the algorithm for the $XYZ$ model. A more detailed discussion will be given elsewhere [22].

We first note that

$$K_0 + K_x \sigma_\mu^x \sigma_\nu^x + K_y \sigma_\mu^y \sigma_\nu^y + K_z \sigma_\mu^z \sigma_\nu^z = \begin{pmatrix} K_1 & 0 & 0 & K_4 \\ 0 & K_2 & K_3 & 0 \\ 0 & K_3 & K_2 & 0 \\ K_4 & 0 & 0 & K_1 \end{pmatrix} \equiv \hat{\Lambda}_{K_1,K_2,K_3,K_4}(\mu,\nu) \quad \text{(A1)}$$

where

$$K_1 \equiv K_0 + K_z, \quad K_2 \equiv K_0 - K_z, \quad K_3 \equiv K_x + K_y, \quad \text{and} \quad K_4 \equiv K_x - K_y. \quad \text{(A2)}$$

Here the constant $K_0$ is irrelevant, i.e., its value does not affect the result at all. We introduced it for appearance. We also define $w_{K_1,K_2,K_3,K_4}$ as the absolute value of the matrix element of $\tilde{\rho}$ which is defined by (3.12) and (3.13) with $\hat{\Lambda}(\mu,\nu)$ replaced by $\hat{\Lambda}_{K_1,K_2,K_3,K_4}(\mu,\nu)$. We then can show, by almost the same argument as we presented to show (3.16), that

$$w_{K_1,K_2,K_3,K_4} = w_{K_1,K_2,-K_3,K_4} = w_{K_1,K_2,K_3,-K_4} = w_{K_1,K_2,-K_3,-K_4}. \quad \text{(A3)}$$

This generalizes (3.16). Because of (A3) we can assume non-negative values for $K_3$ and $K_4$ without loss of generality. We can also assume non-negative $K_1$ and $K_2$ because we can take as large a value as we want for the irrelevant constant $K_0$ without changing final result. Therefore, we assume that all the constants are non-negative.

For the $XYZ$ model, we consider ten types of graphs (Fig. 10) instead of the four types in Fig. 7 for the $XXZ$ model.

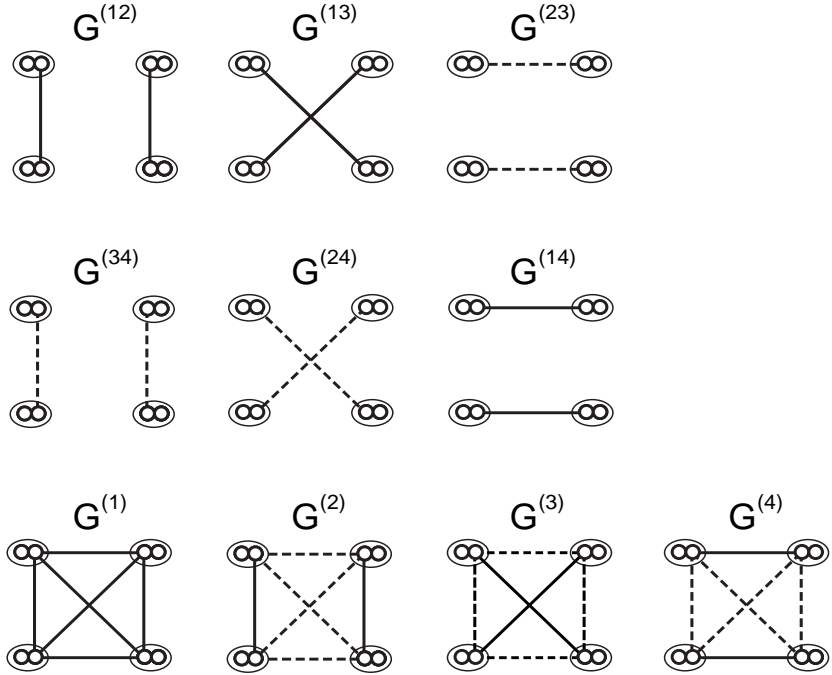

FIG. 10. Ten types of graphs for the decomposition of local Boltzmann factor of the $XYZ$ model. For clarity, two green vertical edges which connect leftmost vertices and rightmost vertices in each graph are not drawn.

We denote these graphs as $G^X(\mu,\nu)$ where $X = (1),(2),(3),(4),(12),(13),(14),(23),(24),$ and $(34)$ as indicated in Fig. 10. Here, we omit the index $p$ specifying a plaquette. We note that the graphs for the four types of operators for the $XXZ$ model discussed can be rewritten in this notation as

$$A(\mu,\nu) = G^{(1,3)}(\mu,\nu),$$
$$B(\mu,\nu) = G^{(2,3)}(\mu,\nu),$$
$$C(\mu,\nu) = G^{(1)}(\mu,\nu),$$
$$D(\mu,\nu) = G^{(2)}(\mu,\nu).$$

The decomposition of the Hamiltonian which is analogous to (4.5), (4.27) and (4.31) is

$$\hat{\Lambda}_{K_1,K_2,K_3,K_4}(\mu,\nu) = \sum_X a_X \hat{\Delta}(G^X(\mu,\nu)), \tag{A4}$$

where the undetermined variables $a_X$ must be non-negative. There is at least one non-trivial solution to this equation. In fact there are many in general. Which solution gives the most effective algorithm has not been studied extensively. The more detailed discussion about the solutions of (A4) will be presented elsewhere.

We should point out here that the matrix element of (A1) can be viewed as the vertex weight of an eight-vertex model with symmetry with respect to simultaneous inversion of all arrows. Therefore, the solution of the equation (A4) gives us a cluster algorithm of an eight-vertex model as well as the foundation for the cluster algorithm of the $XYZ$ model. A similar remark applies to the $XXZ$ model where $S = 1/2$ problem was special case of the six-vertex model.

Here we will only give an relatively simple but useful example of the solution of (A4) instead of listing all possible solutions. We consider the homogeneous system where the coupling constant does not depend on the location. We also assume, without loss of generality, that

$$|K_z| \geq |K_x| \geq |K_y| \geq 0. \tag{A5}$$

(If this is not the case we can 'rotate' the space of spins so that the above inequality holds.) We consider only the case where $K_z > 0$. We can do this without the loss of generality when the original lattice is bipartite. Furthermore, because of (A3) and (A5), we can assume that $K_x \geq 0$ without loss of generality. With these assumptions, we obtain a solution

$$a_1 = K_1 - K_2 - K_3 - K_4 = 2(K_z - K_x),$$
$$a_{(12)} = K_2 = 0,$$
$$a_{(13)} = K_3 = K_x + K_y,$$
$$a_{(14)} = K_4 = K_x - K_y.$$

Correspondingly, (A4) becomes

$$\Lambda_{K_1,K_2,K_3,K_4}(\mu,\nu) = 2(K_z - K_x)\hat{\Delta}(G^{(1)}(\mu,\nu))$$
$$+ (K_x + K_y)\hat{\Delta}(G^{(13)}(\mu,\nu)) + (K_x - K_y)\hat{\Delta}(G^{(14)}(\mu,\nu)). \tag{A6}$$

Here we have chosen $K_0 = K_z$.

In Section IV, we gave an graph-theoretical argument to show the closure of various sets of graphs with respect to multiplication. By a similar argument, we can show that $\Gamma_p^{\text{FXYZ}}$ is closed with respect to multiplication. Here $\Gamma_p^{\text{FXYZ}}$ is defined by

$$\Gamma^{\text{FXYZ}}(p) \equiv \{G \mid V_G = p, \text{ and "All edges in } G \text{ are green."}\}. \tag{A7}$$

We note that

$$\Gamma^{FXYZ}(p) \cap \Gamma^{XXZ}(p) = \Gamma^F(p). \tag{A8}$$

Following the same line of argument as the one in the main text, we can conclude that a set of non-negative variables exists that satisfies

$$\tilde{\rho}_{K_1,K_2,K_3,K_4} = \sum_{G(p) \in \Gamma^{\text{FXYZ}}(p)} v(G(p))\hat{\Delta}(G(p)). \tag{A9}$$

This, of course, leads to a cluster algorithm for the $XYZ$ model with $K_z \geq K_x \geq |K_y| \geq 0$. It is straightforward to calculate $v(G(p))$ analytically for small values of $S$. We note that one useful application of this type of algorithm is the one to the $XY$ model with the $x$-representation basis, i.e., the representation basis in which the $x$-components of spins are diagonalized. This is useful because by this representation we can calculate two-point correlation between in-plane spin components. This correlation may be more interesting to study than correlation between out-of-plane components which can be calculated with the $z$-representation basis.